\documentclass[fleqn,usenatbib]{mnras}

\usepackage{newtxtext,newtxmath}

\usepackage[T1]{fontenc}
\usepackage{ae,aecompl}
\usepackage{tabularx}

\DeclareRobustCommand{\VAN}[3]{#2}
\let\VANthebibliography\thebibliography
\def\thebibliography{\DeclareRobustCommand{\VAN}[3]{##3}\VANthebibliography}

\usepackage{graphicx}	
\usepackage{amsmath}	

\usepackage{amssymb}	

\newcommand{\Msun}{$M_{\odot}$ }

\newcommand{\msun}{M_\odot}

\newcommand{\Romeo}{\texttt{Romeo}}
\newcommand{\Juliet}{\texttt{Juliet}}
\DeclareMathOperator{\sech}{sech}
\newcommand{\tcoolsh}{t_{\rm cool}^{({\rm s})}}
\newcommand{\tff}{t_{\rm ff}}

\title[Thick to thin, bursty to steady]{The bursty origin of the Milky Way thick disc\vspace{-0.1cm}}

\author[S. Yu et al.]{\hspace{-.01cm}Sijie Yu$^{1}$\thanks{E-mail: sijiey3@uci.edu},  
James S. Bullock$^{1}$,
Courtney Klein$^{1}$,
Jonathan Stern$^{2,3}$,
Andrew Wetzel$^{4}$,\newauthor
Xiangcheng Ma$^{5}$,
Jorge Moreno$^{6}$, 
Zachary Hafen$^{1}$,
Alexander B. Gurvich$^{2}$,\newauthor
Philip F. Hopkins$^{7}$,
Du\v{s}an Kere\v{s}$^{8}$,
Claude-Andr\'e Faucher-Gigu\`ere$^{2}$,
Robert Feldmann$^{9}$,\newauthor
Eliot Quataert$^{10}$
\vspace*{5pt} \\
\\
$^{1}$Department of Physics and Astronomy, University of California Irvine, CA 92697, USA \\
$^{2}$Department of Physics \& Astronomy and CIERA, Northwestern University, 1800 Sherman Ave, Evanston, IL 60201, USA \\
$^{3}$School of Physics \& Astronomy, Tel Aviv University, Tel Aviv 69978, Israel \\
$^{4}$Department of Physics and Astronomy, University of California, Davis, CA 95616, USA \\
$^{5}$Department of Astronomy and Theoretical Astrophysics Center, University of California Berkeley, CA 94720, USA \\
$^{6}$Department of Physics and Astronomy, Pomona College, Claremont, CA 95616, USA \\
$^{7}$TAPIR, Mailcode 350-17, California Institute of Technology, Pasadena, CA 91125, USA \\
$^{8}$Department of Physics and Center for Astrophysics and Space Science, University of California at San Diego, 9500 Gilman Drive, La Jolla, CA 92093, USA \\
$^{9}$Institute for Computational Science, University of Zurich, Zurich CH-8067, Switzerland \\
$^{10}$Department of Astrophysical Sciences, Princeton University, Princeton, NJ 08544, USA
}

\date{Accepted XXX. Received YYY; in original form ZZZ}

\pubyear{2021}

\begin{document}
\label{firstpage}
\pagerange{\pageref{firstpage}--\pageref{lastpage}}
\maketitle

\begin{abstract}
We investigate thin and thick stellar disc formation in Milky-Way-mass galaxies using twelve FIRE-2 cosmological zoom-in simulations. All simulated galaxies experience an early period of bursty star formation that transitions to a late-time steady phase of near-constant star formation. 
Stars formed during the late-time steady phase have more circular orbits and thin-disc-like morphology at $z=0$, whilst stars born during the bursty phase have more radial orbits and thick-disc structure.  The median age of thick-disc stars at $z=0$ correlates strongly with this transition time. We also find that galaxies with an earlier transition from bursty to steady star formation have a higher thin-disc fractions at $z=0$.  Three of our systems have minor mergers with LMC-size satellites during the thin-disc phase. These mergers trigger short starbursts but do not destroy the thin disc nor alter broad trends between the star formation transition time and thin/thick disc properties. If our simulations are representative of the Universe, then stellar archaeological studies of the Milky Way (or M31) provide a window into past star-formation modes in the Galaxy. Current age estimates of the Galactic thick disc would suggest that the Milky Way transitioned from bursty to steady phase $\sim$6.5 Gyr ago; prior to that time the Milky Way likely lacked a recognisable thin disc.
\end{abstract}




\section{Introduction}

Milky-Way-mass disc galaxies in the local Universe, including our own, are often characterised by a thin disc component embedded within a thicker disc, which accounts for $\sim$10$-$15\% of total disc luminosity \citep{vdKF2011}. The Milky Way itself has a distribution of disc stars that can be decomposed into thin and thick components spatially \citep{GR83,juric08,Bensby11,BovyRix13}, though there  does not appear to be a bi-modality that defines a clearly distinct thick disc in the Milky Way in terms of age or chemical abundance \citep{Bovy12,Hayden15}.  Whether the there are two distinct components or not, we can use the qualitative terms "thick disc" and "thin disc" as a shorthand classification, one that helps us compare and contrast stars with more eccentric orbits that take them farther from the disc plane to those with more circular orbits that align tightly within it. 

In the Milky Way today, the thin disc is home to recent star formation.  Thicker-disc stars tend to be older, more metal poor, and more alpha-enhanced  \citep{Furhmann88,Haywood13,RB14,Hayden15,Mackereth17,Hayden17,Sharma19}. These characteristics may be loosely interpreted as evidence that thick disc stars formed early and rapidly, perhaps in a series of bursts \citep[e.g.][]{vdKF2011}. \citet{Snaith14} use elemental abundances of long-lived stars to derive a star-formation history for a selection of old, alpha-enhanced stars they associate with the Milky Way thick disc. They conclude that these stars emerged during an early, elevated period of star formation -- and that the Galactic thick disc may be comparable {\em in mass} (not luminosity) to the young (< 8 Gyr old) thin disc. 

Despite years of study, an understanding of how thick and thin discs arise within the broader story of galaxy formation remains a key question. One of the most enduring ideas is that pre-existing thin discs are heated in mergers with small satellite galaxies to create a vertically extended component \citep{Quinn93,Kazantzidis08,Purcell09}. In fact, the phase-space structure of stars in the solar neighborhood provides some evidence that such a event -- the Gaia-Enceladus Sausage merger -- may have been significant enough to heat a proto Milky Way disc, under the assumption that a thin disc existed at this early time \citep{Belokurov18,Helmi18}. However, as hinted above, the chemical abundance structure of the Milky Way disfavours the idea that thick-disc formation is associated with a single discrete event \citep{Freundenburg17}. Rather, these data favour an ``upside down'' formation scenario -- first predicted by cosmological simulations \citep{Brook04,Brook06,Brook12,Bird13} -- where stars are born in relatively thick discs at early times, and only later form in thin discs.

Many recent cosmological simulations naturally produce $z=0$ discs with young stars concentrated in a thinner component than old stars \citep[e.g.][]{Ma2017,Navarro18,Grand18,Pillepich19,Bird20,Park20}. These same simulations
at high redshift produce discs that are systematically thicker and clumpier than those at low redshift, as observed in nearly all deep, high-resolution imaging studies of galaxies \citep{Elmegreen06,Elmegreen07,Shapiro2008,Genzel2008,Overzier2010,Elmegreen17}. The observed transition from thick irregular galaxies at high redshift to thin rotation-dominated discs at low redshift is well established, and often referred to as ``disc settling'' \citep{Kassin12}. 


\begin{table*}
  \caption{The simulations we employ in this work. We list the following: the name of the zoom-in target halo, the stellar mass ($M_{\star}$) within the central 20 kpc of the halo at $z=0$, the radius (${R_\mathrm{90}}$) enclosing 90\% of $M_{\star}$, the halo virial mass ($M_{\mathrm{halo}}$), the halo virial radius ($R_{\mathrm{halo}}$), the resolution of each simulation quantified by the initial baryonic particle mass ($m_{\mathrm{i}}$), and the reference that first introduced each halo at the quoted targeted resolution. The remaining columns present derived quantities: the lookback time to the end of the bursty phase/onset of the steady phase ($t_{\rm B}$), the mass-weighted thin-disc fraction ($f_{\rm thin \, disc \, m}$), the luminosity-weighted thin-disc fraction ($f_{\rm thin \, disc \, l}$), the median thick-disc age ($t_{\rm 50}$), and the 90\% oldest star of the thick disc ($t_{\rm 90}$). Hosts with names starting with `m12' are isolated configurations selected from the Latte suite, whilst the rest are in LG-like pairs from the ELVIS on FIRE suite. The four galaxies marked with an asterisk correspond to short-lived, late-time bursts of star formation taking place after the onset of the steady phase. Three of these four bursts appear to be triggered by minor mergers. The exception is \texttt{Thelma}, which has late-time star formation in the ``steady'' regime (by our formal definition), but is still experiencing fairly variable star formation compared to most of our galaxies at late times. These bursts and/or mergers tend to influence $t_{\rm 90}$ but do not significantly affect $t_{\rm 50}$ nor $f_{\rm thin\ disc}$. The haloes are ordered by decreasing $t_{\rm B}$, from $t_{\rm B}$ = 6.52 Gyr (\texttt{Romeo}, top) to $t_{\rm B}$ = 0.0 Gyr (\texttt{m12w}, bottom). } 
	\centering 
	\label{tab:info}
	\begin{tabularx}{\textwidth}{Xccccccccccc}
		\hline
		\hline  
		Simulation & $M_{\star}$ & ${R_\mathrm{90}}$ &  $M_{\mathrm{halo}}$ & $R_{\mathrm{halo}}$ & $m_{\mathrm{i}}$ & $t_{\mathrm{B}}$ & $f_{\mathrm{thin\ disc\ m}}$ & $f_{\mathrm{thin\ disc\ l}}$ & thick disc $t_{50}$ & thick disc $t_{90}$ & Reference \\
		Name  &  $[M_{\odot}]$ & [kpc] & $[M_{\odot}]$ & [kpc] & $[M_{\odot}]$ & [Gyr] & ($M$ weighted) & ($L$ weighted) & [Gyr] & [Gyr] \\
		\hline 
		\texttt{Romeo} & 7.4$\times$10$^{10}$ & 13.3 & 1.0$\times$10$^{12}$ & 317 & 3500 & 6.52 & 0.45 & 0.70 & 8.96 & 6.16 & A\\
		\texttt{m12b}* & 8.1$\times$10$^{10}$ & 9.8 & 1.1$\times$10$^{12}$ & 335 & 7070 & 6.32 & 0.37 & 0.64 & 7.34 & 2.72 & A\\
		\texttt{Remus} & 5.1$\times$10$^{10}$ & 12.3 & 9.7$\times$10$^{12}$ & 320 & 4000 & 5.88 &0.36 & 0.62 & 8.22 & 4.88 & B\\
		\texttt{Louise} & 2.9$\times$10$^{10}$ & 12.0 & 8.5$\times$10$^{11}$ & 310 & 4000 & 5.56 & 0.32 & 0.65 & 8.11 & 4.06 & A\\
		\texttt{m12f}* & 8.6$\times$10$^{10}$ & 11.0 & 1.3$\times$10$^{12}$ & 355 & 7070 & 5.01 & 0.38 & 0.65 & 6.28 & 2.62 & C\\
		\texttt{Romulus} & 1.0$\times$10$^{11}$ & 14.2 & 1.5$\times$10$^{12}$ & 375 & 4000 & 4.90 & 0.37 & 0.69 & 7.37 & 4.92 & B\\
		\texttt{Juliet} & 4.2$\times$10$^{10}$ & 9.6 & 8.5$\times$10$^{11}$ & 302 & 3500 & 4.40 & 0.30 & 0.62 & 6.74 & 4.66 & A\\
		\texttt{m12m} & 1.1$\times$10$^{11}$ & 11.3 & 1.2$\times$10$^{12}$ & 342 & 7070 & 3.81 & 0.34 & 0.58 & 6.07 & 3.24 & E\\
		\texttt{m12c}* & 6.0$\times$10$^{10}$ & 9.7 & 1.1$\times$10$^{12}$ & 328 & 7070 & 3.70 & 0.32 & 0.62 & 5.39 & 2.30 & A\\
		\texttt{m12i} & 6.4$\times$10$^{10}$ & 9.2 & 9.0$\times$10$^{11}$ & 314 & 7070 & 3.14 & 0.32 & 0.59 & 6.18 & 3.50 & D\\
		\texttt{Thelma}* & 7.9$\times$10$^{10}$ & 12.4 & 1.1$\times$10$^{12}$ & 332 & 4000 & 2.57 & 0.27 & 0.57 & 4.73 & 1.95 & A\\
        \texttt{m12w} & 5.8$\times$10$^{10}$ & 8.7 & 8.3$\times$10$^{11}$ & 301 & 7070 & 0.0 & 0.24 & 0.43 & 4.38 & 1.13 & F\\
		\hline  
	\end{tabularx}
	\raggedright
	\textit{Note}: The references are: 
	A:~\cite{Garrison-Kimmel19},
	B:~\cite{Garrison-Kimmel19_2}
	C:~\cite{Garrison-Kimmel17}, 
	D:~\cite{Wetzel16}, 
	E:~\cite{Hopkins17}, 
	F:~\cite{Samuel20}.
	\label{tab:one}
\end{table*}

Whilst upside-down disc formation is seen regularly in simulations, the physical origin of this thick-to-thin transition has been hard to distill. One idea is that discs are born thick during an early period of gas-rich mergers \citep{Brook04}. At high redshift, high star-formation rate densities, high gas fractions, and feedback-induced turbulence can also contribute to an initially hot disc \citep{Lenhert14}. An alternative possibility is that stars are initially born in thin discs, but are quickly and continuously heated owing to chaotic accretion of gas \citep{Meng2020}. In some simulations, most stars are born kinematically hotter at early times {\em and} subsequently heated after birth on a short timescale \citep{Ma2017,Bird20}. 

In this paper, we explore the transition from thick to thin disc formation in twelve Milky-way-mass galaxies drawn from FIRE-2 cosmological zoom-in simulations. As seen in previous work \citep[][]{Brook04,Bird13,Ma2017,Navarro18,Bird20,Park20}, our discs tend to form upside down, with the thick discs in place early and the thin disc forming at late times. One new finding in our work is that the transition from thick to thin-disc formation correlates with a transition in star formation mode, from an early, elevated bursty phase with highly time-variable star formation rate, to a late-time steady phase of near-constant star formation rate. Thin-disc stars tend to be born during the late-time steady phase, whilst thick-disc stars are associated with the latter part of the bursty phase.  Galaxies with older thick-disc populations have an earlier transition from bursty to steady star formation. The earlier the transition time, the more dominant the thin disc is at $z=0$.  

A transition from bursty to steady star formation has been reported previously in the FIRE simulations, at $z = 0.5 - 1.5$ in Milky-Way-mass haloes \citep{Muratov15,Sparre17,Angles2017,AA17,FG18}. In particular, \citet{Stern20} show that the transition to steady star formation coincides with the virialisation of the inner circumgalactic medium (CGM). Specifically, when haloes in FIRE cross a characteristic mass scale ($\sim$ $10^{12}$ M$_\odot$), the cooling time of shocked gas in the inner halo ($0.1$ R$_{\rm vir}$) exceeds the local free-fall time. This creates a hot confining medium, with high and nearly uniform thermal pressure. After this time, \citet{Stern20} observes that star formation becomes less bursty and gaseous discs become more rotationally supported. This steady, settled disc phase may be enabled by the hot, pressurised inner CGM itself, which may prevent supernova-driven outflows from repeatedly blowing out the interstellar medium (ISM) in a way that would otherwise might perturb disc structure \citep[e.g.][]{Martizzi20}. 

Of particular relevance is work by \citet{Ma2017}, who used a slightly lower resolution FIRE-1 zoom-in simulation to show that disc stars at large scale-height (thick disc stars) form primarily during an early chaotic bursty mode, whilst younger stars were formed in a more stable disc. In what follows, we perform a more systematic analysis using a larger, higher resolution sample of FIRE-2 haloes and confirm that this result is more general. This motivates us to suggest that the physical transition from bursty to steady star formation also coincides with a shift from thick-disc to thin-disc formation in Milky-Way-mass galaxies. If this is true in the real Universe, then stellar archaeological studies of the Milky Way could provide a window into past star-formation modes, as well as the build-up of the Galactic CGM. 

Below we elect to define thick and thin disc populations using a purely kinematic definition based on each star particle's circularity \citep[][]{Abadi03}. Whilst it is common in Milky Way studies to use elemental abundances to divide thin and thick disc populations, we adopt this kinematic definition in order to fully decouple our selection from the nature of star formation. Specifically, alpha enhancement correlates with starburst activities, so we would like to avoid using abundance ratios when looking for distinct correlations related to star formation history. The fact that we find correlations between kinetically-defined thick-disc populations and the mode of star formation suggests that the correlation between thick disc formation and star formation activity is non-trivial. Moreover, the previous work by \citet{Ma2017} finds qualitatively similar results using more traditional observationally-oriented definitions of the thick disc based on a vertical density profile, suggesting that the result is insensitive to selection choices.

The outline of this manuscript is as follows.  In Section \ref{sec:sims} we provide an overview of our simulations and present our kinematic definition of thin and thick disc stars.  Section \ref{sec:results} presents results focusing for two illustrative cases (\ref{sec:examples}) and then on to explore general trends for all galaxies in our sample (\ref{sec:sample-wide}).  We conclude and discuss our results in the context of the Milky Way in Section \ref{sec:conclusions}.

\section{Simulations and methods}
\label{sec:sims}

\subsection{FIRE-2 simulations of Milky-Way-mass galaxies}
\label{sec:sims_intro}

Our analysis utilises cosmological zoom-in simulations performed with the multi-method gravity plus hydrodynamics code {\small GIZMO} \citep{Hopkins15} from the Feedback In Realistic Environments (FIRE) project\footnote{\url{https://fire.northwestern.edu/}}. We rely on the FIRE-2 feedback implementation \citep{Hopkins17} and the mesh-free Lagrangian Godunov (MFM) method. The MFM approach provides adaptive spatial resolution and maintains conservation of mass, energy, and momentum. FIRE-2 includes radiative heating and cooling for gas across a temperature range of $10-10^{10}$K. Heating sources include an ionising background \citep{Faucher2009}, stellar feedback from OB stars, AGB mass-loss, type Ia and type II supernovae, photoelectric heating, and radiation pressure -- with inputs taken directly from stellar evolution models. The simulations self-consistently generate and track 11 elemental abundances (H, HE, C, N, O, Ne, Mg, Si, S, Ca, and Fe), and include sub-grid diffusion of these elements in gas via turbulence \citep{Hopkins2016,Su17,Escala18}. Star formation occurs in gas that is locally self-gravitating, sufficiently dense ($ > 1000$ cm$^{-3}$), Jeans unstable and molecular (following \citealt{Krumholz_2011}). Locally, star formation efficiency is set to $100\%$ per free-fall time; i.e., $\rm SFR_{\rm particle} = m_{\rm particle} \cdot f_{\rm mol} \, / \, t_{\rm ff}$. Gas particles are converted to stars at this rate stochastically \citep{Katz1996}. Note that this does {\em not} imply that the global efficiency of star formation is $100\%$ within a giant-molecular cloud (or across larger scales). Self-regulated feedback limits star formation to $\sim$1-10\% per free-fall time \citep{Faucher2009,Hopkins17_2,Orr_2018}.

In this work, we analyse 12 Milky-Way-mass galaxies (Table \ref{tab:one}). These zoom simulations are initialised following \citet{Onorbe14}. Six of these galaxies (with names following the convention \texttt{m12*}) are isolated and part of the Latte suite ~\citep{Wetzel16,Garrison-Kimmel17,Hopkins17_2,Garrison-Kimmel19}. Six, with names associated with famous duos (e.g. \Romeo~and \Juliet), are part of the ELVIS on FIRE project \citep{Garrison-Kimmel19,Garrison-Kimmel19_2} and are set in Local-Group-like configurations, as in the ELVIS suite \citep{Garrison-Kimmel14}. This suite includes three simulations, containing two MW/M31-mass galaxies each. The main haloes were selected so that they have similar relative separations and velocities as of the MW-M31 pair in the Local Group (LG).   Table \ref{tab:one} lists the initial baryonic particle masses for each simulation. Latte gas and star particles have initial masses of $7070\, \msun$, whilst ELVIS on FIRE has $\approx 2 \times$ better mass resolution ($m_{\rm i} = 3500 - 4000\, \msun$). Gas softening lengths are fully adaptive down to $\simeq$0.5$-$1 pc. Star particle softening lengths are $\simeq$4 pc physical and a dark matter force softening is $\simeq$40 pc physical.

\begin{figure*}
	\includegraphics[width=0.94 \textwidth, trim = 30.0 0.0 30.0 0.0]{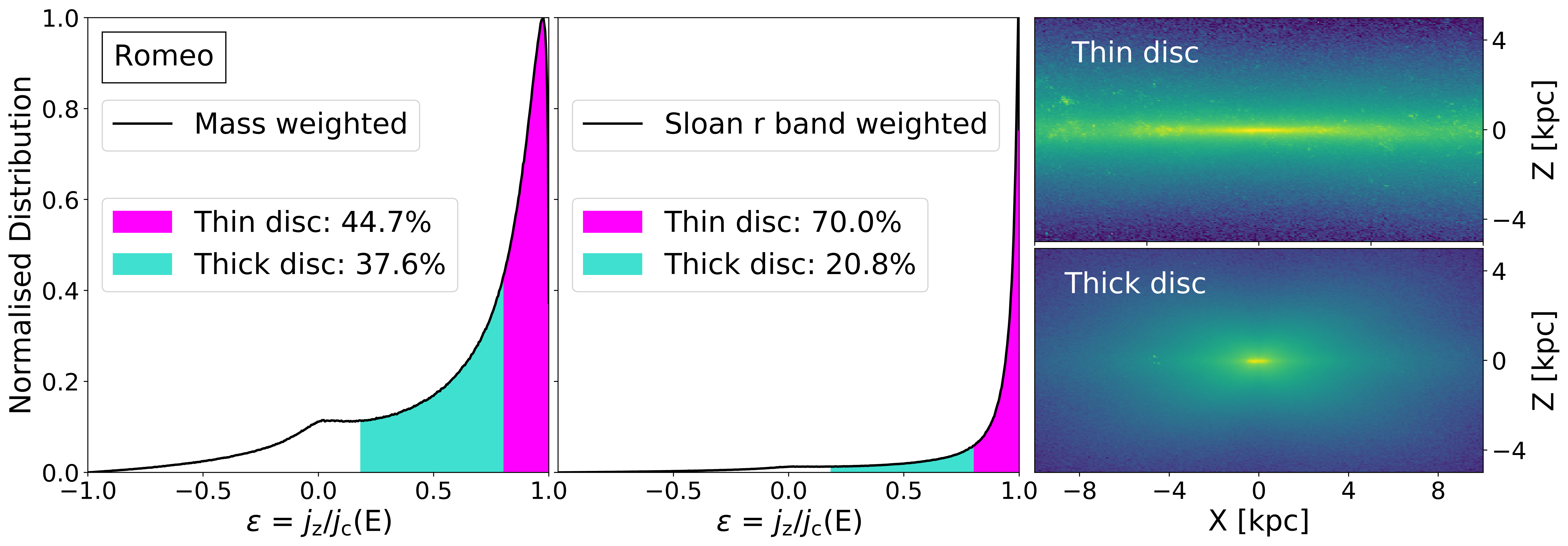}
	\includegraphics[width=0.94 \textwidth, trim = 30.0 0.0 30.0 0.0]{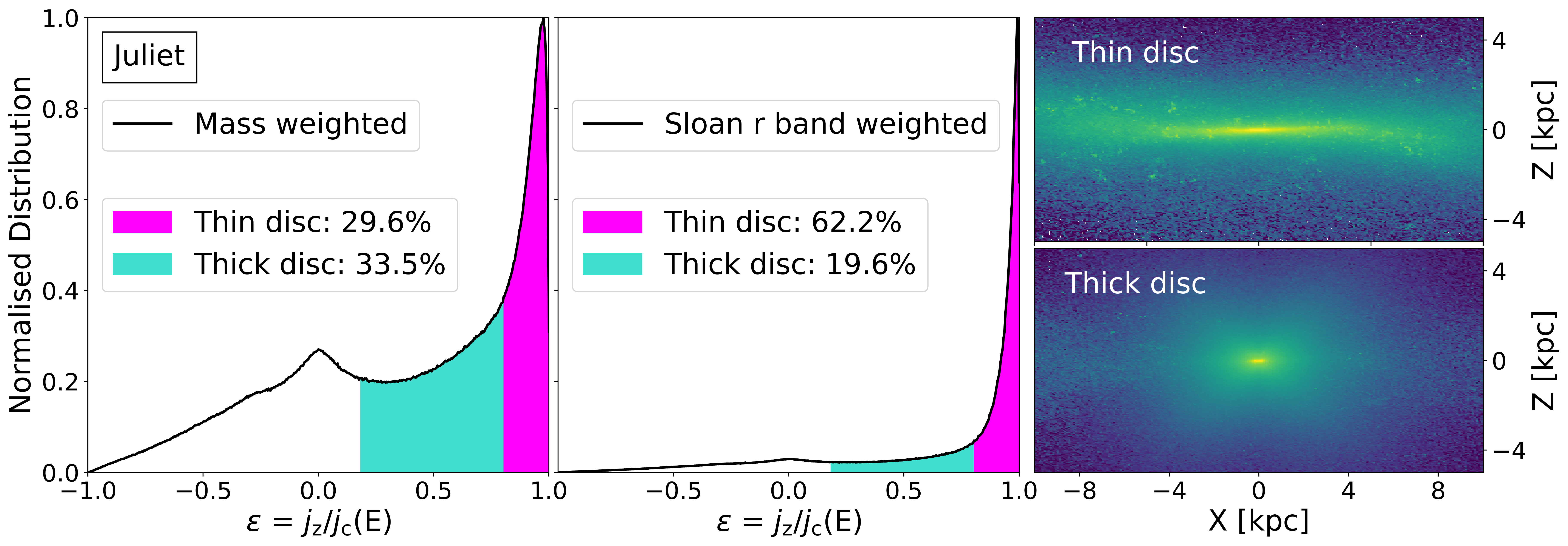}
    \caption[]{Circularity and spatial distributions of stars in \texttt{Romeo} (top set) and \texttt{Juliet} (bottom set). The {\bf left panels} show the mass-weighted distribution of circularities $\epsilon$ of all the stars within $R_{90}$ in the galaxies ($13.3$ kpc and $9.6$ kpc, respectively). See \S \ref{sec:define} for a description of $\epsilon$. The magenta blocks mark thin disc stars, which we define as the stars with $\epsilon\geq 0.8$. The cyan blocks mark thick disc stars, which we define to be those with $0.8 >\epsilon\geq 0.2$. The fraction of stars in each block is shown in the legend. The {\bf middle panels} show the same distributions but now weighted by the Sloan \textit{r} band luminosity of each star particle. The luminosity-weighted distributions generally give a much higher thin-disc fraction. The percentages indicate mass- and luminosity-weighted fractions for each component. The {\bf right panels} display $z=0$ edge-on views (2D density weighted by Sloan \textit{r} band luminosity) of the thin (upper) and thick (lower) disc stars. We see that these definitions produce disc components that qualitatively resemble geometrically-defined thin and thick discs. }
	\label{fig:circ_dist+spatial}
\end{figure*}

\subsection{Defining thin and thick discs}
\label{sec:define}

\begin{figure*}
	\includegraphics[width=0.999 \textwidth, trim = 0.0 0.0 0.0 0.0]{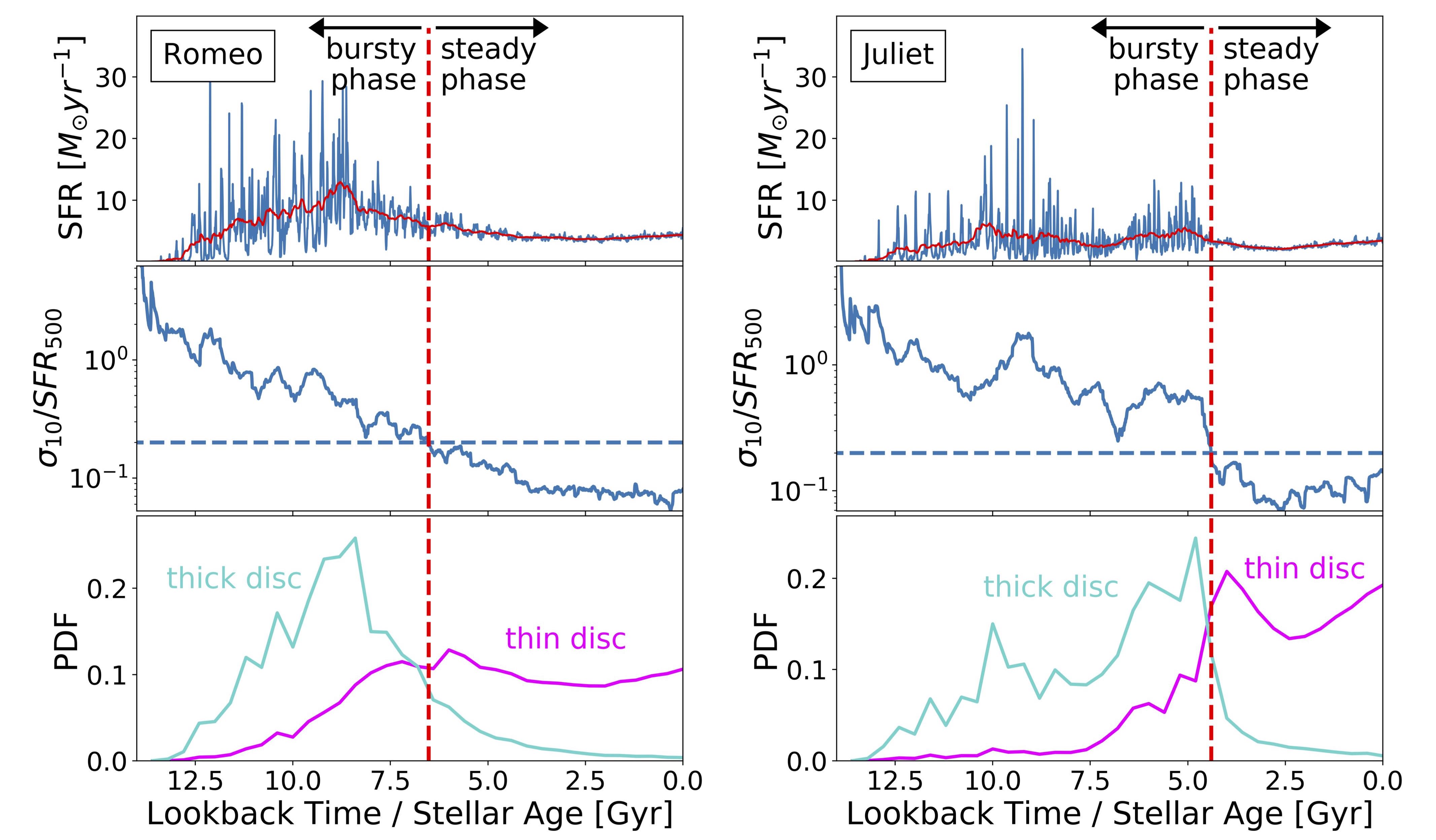}
    \caption[]{ Star formation histories and thin/thick disc stellar age distributions for \texttt{Romeo} (left) and \texttt{Juliet} (right). The {\bf top panels} show the star formation rate in the galaxy as a function of lookback time. The 
    blue lines show the ``instantaneous'' star formation rate averaged over 10 Myr bins, while the red lines show the ``smoothed'' star formation rate averaged over 500 Myr bins.
    The {\bf middle panels} shows the variance in instantaneous star formation rate divided by the smoothed star formation rate as a function of time.  
    We see that the SFR variance in each galaxy generally decreases with time, from bursty to steady, as we approach the present day. For ease of description, we  divide the star formation history of each galaxy into two distinct phases -- an early bursty phase and a late steady phase -- delineated a time $t_{\rm B}$ where the SFR variance falls below $0.2$ times the smoothed star formation rate.  This "bursty-phase lookback time" is marked by the vertical red dashed line in each upper panel. The {\bf bottom panels} shows the age distribution of $z=0$ stars that belong to the thick disc (cyan) and thin disc (magenta) in each galaxy.  We see that thick-disc stars have ages that track closely the bursty period of star formation while thin-disc stellar ages correspond more closely to the steady phase.}
	\label{fig:SFH_Age_distribution_RJ}
\end{figure*}

There are multiple ways to separate a ``thick disc'' from a ``thin disc'' population in observations \citep{Martig16}.  The physical characteristics authors use to define the thick disc include geometric morphology, kinematics, chemical abundances, and age. The geometric/morphological definition is the natural choice for distant galaxies, where detailed chemical and/or age information is harder to extract. In this theoretical analysis, we adopt a simple kinematic definition that allows us to identify thin and thick disc populations cleanly, and which also produces disc populations that follow the qualitative geometric expectations for thin and thick discs. We elect to avoid definitions based on chemical-enrichment history because we are exploring correlations with star-formation burstiness, and we opt to avoid any selection effects such a choice might impose. 

Our categorisation is based on each star particle's circularity, $\epsilon = j_z/j_c(E)$, defined as the ratio of each particle's angular momentum to that of a circular orbit with the same energy \citep{Abadi03}. The angular momentum direction $\hat{z}$ is set by total stellar angular momentum within $10$ kpc of each galaxy's center. 
We categorise star particles with $\epsilon = 0.8 - 1$ as {\it thin disc} stars, and those with $\epsilon = 0.2 - 0.8$ as {\it thick disc} stars. Our classification is motivated by past explorations \citep[e.g.][]{Abadi03,Okamoto2010,Knebe2013}, which find that circularity correlates well with standard morphological definitions of thin and thick disc populations.

The left and middle panels of Figure \ref{fig:circ_dist+spatial} illustrate our circularity-based definitions for two specific simulations: \Romeo~(top) and \Juliet~(bottom). The left panels show the mass-weighted circularity distributions and the middle panels show the luminosity-weighted circularity distributions for each galaxy. By our definitions, the magenta regions correspond to thin-disc stars, whilst the cyan regions correspond to thick-disc stars. Note that whilst the mass-weighted distributions yield approximately equal thin and thick disc populations, the luminosity-weighted distributions assign $\sim$60$-$70\% to the thin disc. The right panels show luminosity-weighted images of the thin and thick disc populations for each galaxy, which illustrate that our orbit-based definitions yield spatial distributions that look qualitatively like discs that are indeed thin and thick.

We find that our $\epsilon$-based classification scheme results in thin and thick disc populations with vertical density profiles (in the $z$ direction) resembling those of traditional morphologically identified thin and thick discs. Whilst some of our galaxies have vertical profiles better fit by exponential forms, the majority prefers $\sech^2$ fits. At mock solar locations (8 kpc from the galactic centre), fits to the resultant thin-disc populations yield scale heights for our 12 galaxies that range from $\sim$250 pc to $\sim$800 pc for luminosity-density profiles (in Sloan \textit{r} band); and from $\sim$500 pc to $\sim$950 pc for mass-density profiles. Similar fits to our thick-disc populations have scale heights that range from $\sim$1.2$-$1.5 kpc for luminosity-density profiles. These results are consistent with previous analysis \citep[e.g.][]{Ma2017,Sanderson2020}. We find that dividing populations in this manner yields scale-height results in line with those we obtain with more traditional (purely spatially-based) two-component fits.  We also find that our simple $\epsilon$-based classification yields thick disc populations that are older, more metal poor, and more alpha enhanced than the thin discs we identify.

We note that there can be a non-trivial fraction of stars that exist at very low or negative circularities ($\epsilon < 0.2$), which would naturally be associated with a spheroidal component. 
For example, in Figure \ref{fig:circ_dist+spatial}, for \Romeo \ (\Juliet), this component represents 17\% (37\%) of the mass and 6.8\% (13\%) of the light. We generally find that these spheroidal stars tend to form in the earliest periods of galaxy assembly, whilst thick-disc stars form later. Since the focus of this paper is on thin/thick disc formation, we have largely ignored low/negative angular momentum stars in what follows, though further investigation into the origin of the inner spheroid as it relates to star formation in the early galaxy is warranted. Such an exploration would require a more sophisticated kinematic disc/spheroidal classifications of stars with overlapping $\epsilon$ ranges.  We have performed a simple check of the sensitivity of our main results to the presence of bulge stars by neglecting all stars that sit within 1 kpc of the galactic center of each galaxy and find that this does not change our results substantially. The fraction of stars that have $\epsilon > 0.2$ and that sit within 1 kpc is relatively small in all of our galaxies and, when either excluded or included, have only a minor effect on the age distributions of our "thick disc" stars.

\section{Results}
\label{sec:results}

\subsection{Two Illustrative Cases: \Romeo~and \Juliet}
\label{sec:examples}

\subsubsection{Bursty phase, steady phase, and age distributions}

The top panels in Figure \ref{fig:SFH_Age_distribution_RJ} show the star formation histories of \Romeo~(left) and \Juliet~(right) as a function of lookback time. The star formation rate~\footnote{These star formation histories are measured for all particles that were born within 20 kpc of the most massive progenitor.}  (SFR) displayed is averaged over both a short timescale of $10$ Myr (SFR$_{10}$, blue) and a longer timescale of $500$ Myr (SFR$_{500}$, red). 
 The middle panel shows the variance in ``instantaneous'' SFR,
 $\sigma_{10}$, divided by the average SFR$_{500}$ as a function of lookback time. We define $\sigma_{10}(t)$ as the variance in SFR$_{10}$ over a time range spanning $t \pm 250$ Myr. 
 We see that the relative variance is much larger at early times than at late times, consistent with previous studies \citep[e.g.][]{Stern20,Jose2020} that have shown that star formation in massive FIRE-2 galaxies tends to transition from bursty to steady as we approach the present day. 
 
While the transition from bursty to steady is not always sharp, the trend is quasi-monotonic, with the the ratio $\sigma_{10}/$SFR$_{500}$ generally decreasing with time. For the sake of simplicity in this analysis, we find it useful to  divide the star formation history of each galaxy into two distinct phases: an early bursty phase and a late-time steady phase. We define the bursty phase to end at a lookback time $t_{\rm B}$ when the variance in ``instantaneous'' star-formation rate  first falls below $B=0.2$ times the time-averaged star formation rate:
\begin{equation}
  \frac{\sigma_{10}(t_{\rm B})}{{\rm SFR}_{500}(t_{\rm B})} \equiv B.
  \label{eq:def}
\end{equation}
We use this definition to assign a specific bursty-phase timescale to each galaxy's star formation history.  Our qualitative results are not sensitive to the precise choice of $B = 0.2$ on the right-hand side of Equation \ref{eq:def}. Larger choices ($B>0.2$) tend to push the bursty phase slightly earlier and smaller choices ($B<0.2$) tend to push the burst phase slightly later.  By our adopted definition,  \Romeo~ has a bursty-phase lookback time of $t_{\rm B}= 6.5$ Gyr and \Juliet~ has a bursty phase that ends more recently at $t_{\rm B} = 4.4$ Gyr. The vertical, red-dashed lines in Figure \ref{fig:SFH_Age_distribution_RJ} mark these times. Table \ref{tab:one} provides bursty-phase lookback times $t_{\rm B}$ for our simulated galaxies.

The bottom panels in Figure  \ref{fig:SFH_Age_distribution_RJ} show the age distributions of thick-disc stars (cyan) and thin-disc stars (magenta). Thick-disc ages tend to track the bursty-phase star formation, whilst the thin disc stars closely track the steady phase in each case. We emphasize again that in defining a specific value for $t_{\rm B}$ we do not mean to suggest that there is always a razor-sharp phase-change in star formation activity (or in disc thickness) but rather to assign a specific timescale to each galaxy that reasonably marks a qualitative transition.  We note that age-overlap of thick and thin disc stars in \Romeo~ is much more significant than it is in \Juliet.  This mirrors the more gradual decrease in relative SFR variance in \Romeo, compared to the sharp transition near $t_{\rm B}$ seen in \Juliet.
Nevertheless, the broad tendency for typical thick disc stellar ages to correlate with bursty-phase lookback times is seen for every galaxy in our sample (as we show in Section~\ref{sec:sample-wide} below).

\subsubsection{Morphology with time}

Figure \ref{fig:Romeo_morph_compare} and \ref{fig:Juliet_morph_compare} show images of \Romeo~ and \Juliet~ at three specific times in the past: 8.4, 4.7, and 2.7 Gyr ago, which also illustrate how stars that formed at these epochs are spatially distributed today. The top row (a) shows the star formation rate versus time. The arrow symbols on the time axis indicate the specific lookback times visualized beneath. Row (b) shows luminosity-weighted images of the main progenitor of each galaxy at the specified times.  Each snapshot is viewed edge-on with respect to the stellar angular momentum axis at that time.  Row (c) includes images of the young stellar populations, corresponding to stars born within the last 100 Myr of the indicated times. Lastly, row (d) shows the current location ($z=0$) of the young stars shown in (c).  Note that rows (c) and (d) are similar to  Figure 1 in \citet{Ma2017}.
   
Figure \ref{fig:Romeo_morph_compare} shows that, $8.4$ Gyr ago (prior to the end of the bursty phase) \Romeo~ resembled a thick disc embedded within a significant spheroid. The stars forming at this time (panel c, far left) are not very well ordered into a thin disc, but do exhibit some coherence. Those stars today are arranged in a thick-disc like configuration (d, far left). Conversely, at $4.7$ Gyr and $2.7$ Gyr (after the steady phase has commenced) \Romeo's thin disc has fully emerged. Young stars at those times are situated in thin discs (c, middle and right) and remain in relatively thin configurations at $z=0$  (d, middle and right).

\begin{figure*}
	\includegraphics[width=0.99 \textwidth, trim = 280 0.0 400 0.0]{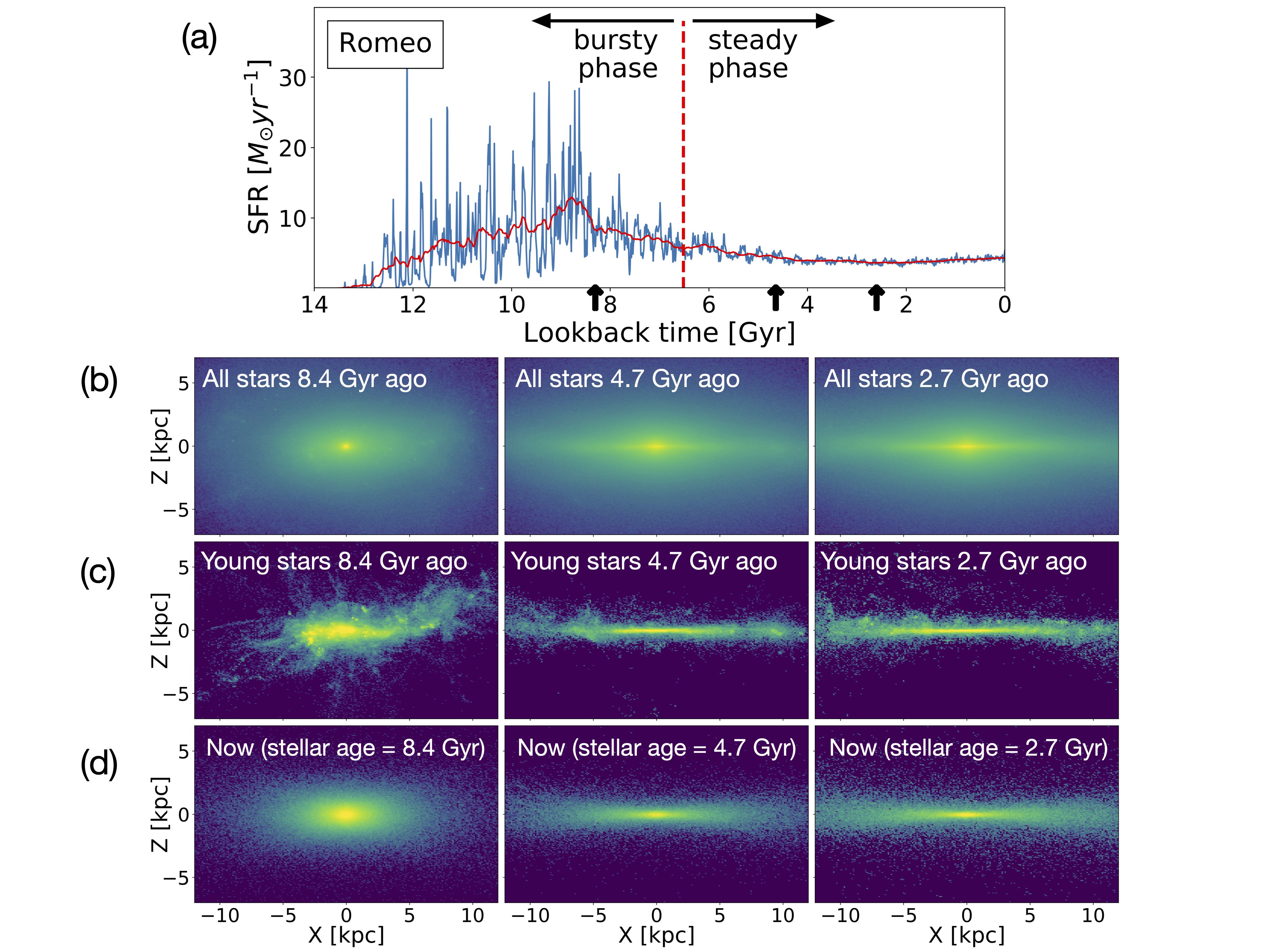}
    \caption[]{ (a) Instantaneous (blue) and smoothed (red) star formation rate for \texttt{Romeo} plotted as a function of lookback time.  The vertical red dashed line shows our adopted time that separates the bursty phase from the steady phase according to Equation \ref{eq:def}. (b) Edge-on, luminosity-weighted images for \texttt{Romeo} at three different lookback times -- from left to right: 8.4 Gyr, 4.7 Gyr, and 2.7 Gyr.  The arrows along the time axis in top panel indicate these times. (c) Edge-on, luminosity-weighted images for the youngest population (formed within 100 Myr) at the same times. (d) Edge-on, luminosity-weighted images at $z = 0$ for the same stars shown in row c. Note that 8.4 Gyr ago, when the star formation was still bursty, the galaxy resembles a thick disc. At the later two times, in the steady phase, a thin-disc component emerges. 
 }
	\label{fig:Romeo_morph_compare}
\end{figure*}

\begin{figure*}
	\includegraphics[width=0.95 \textwidth, trim = 280 0.0 400 0.0]{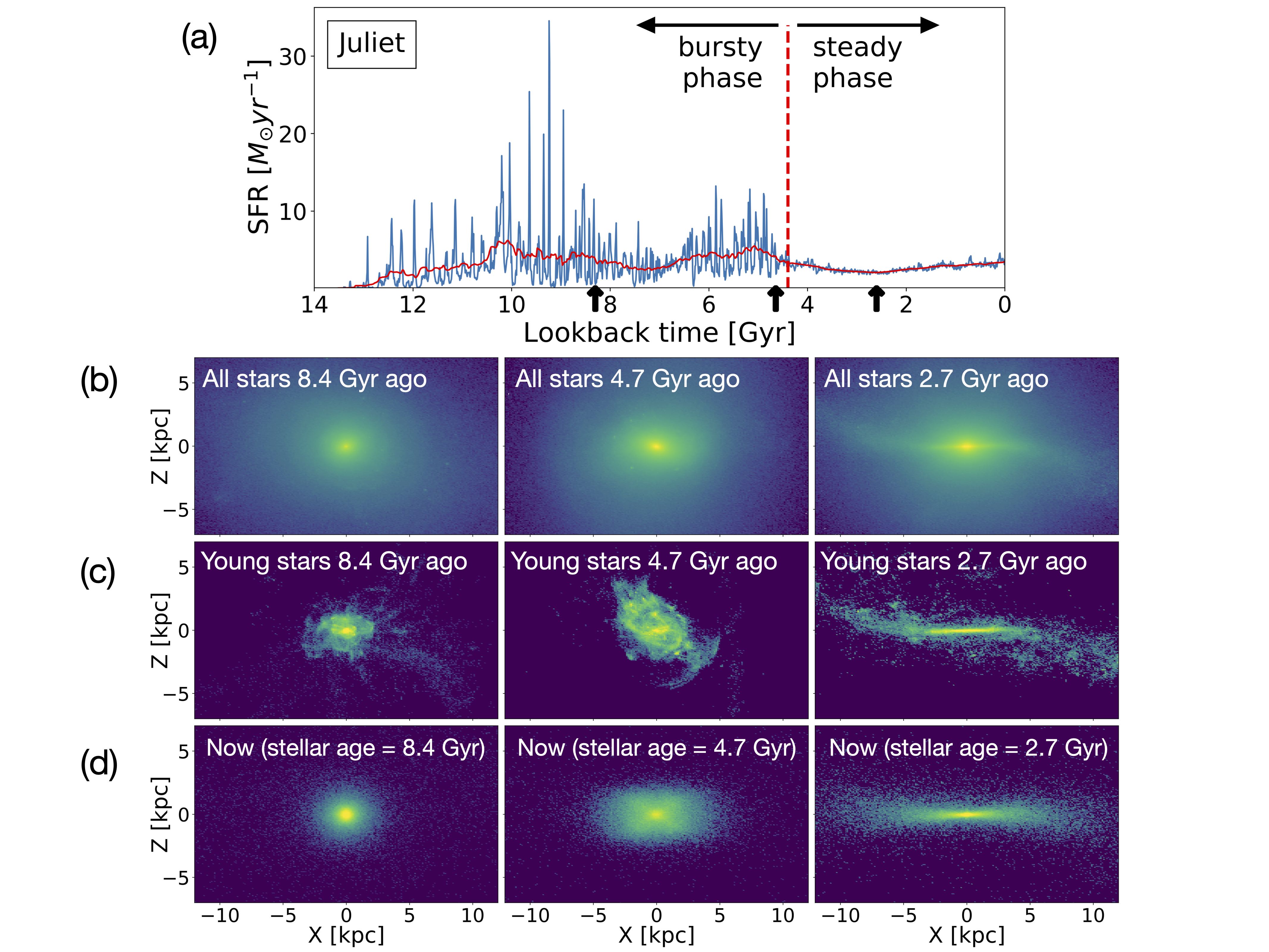}
    \caption[]{Same as Figure \ref{fig:Romeo_morph_compare}, now for  \texttt{Juliet}. At 8.4 Gyr and 4.7 Gyr ago, when \Juliet's star formation was still bursty, there is no visible thin disc.  Only in the $2.7$ Gyr images, after the steady phase has started, is thin-disc morphology apparent.}
	\label{fig:Juliet_morph_compare}
\end{figure*}

Figure \ref{fig:Juliet_morph_compare} shows that \Juliet~ exhibits a transition from thick to thin, which happens later than \Romeo's. Concretely, whilst \Romeo~had a pronounced thin disc component $4.7$ Gyr ago, \Juliet~ had no thin disc at that time. Only in the most recent image (2.7 Gyr) does \Juliet~ begin to resemble a thin disc.  This difference in morphological structure with time mirrors the difference we see in the transition to steady star formation.  \Juliet~ has a bursty phase that ends only at a lookback time of $t_{\rm B} = 4.2$ Gyr, compared to \Romeo, which ended its bursty phase $t_{\rm B}= 6.7$ Gyr ago. At $4.2$ Gyr, \Juliet~ happens to have just experienced a rapid inflow of cool gas, some of which has formed stars in the thick, rotating structure we see in row (c), middle panel. Those stars end up in a thick disc component at $z=0$ (row d, bottom).

\begin{figure*}
	\includegraphics[width=0.95 \textwidth, trim = 30.0 0.0 30.0 0.0]{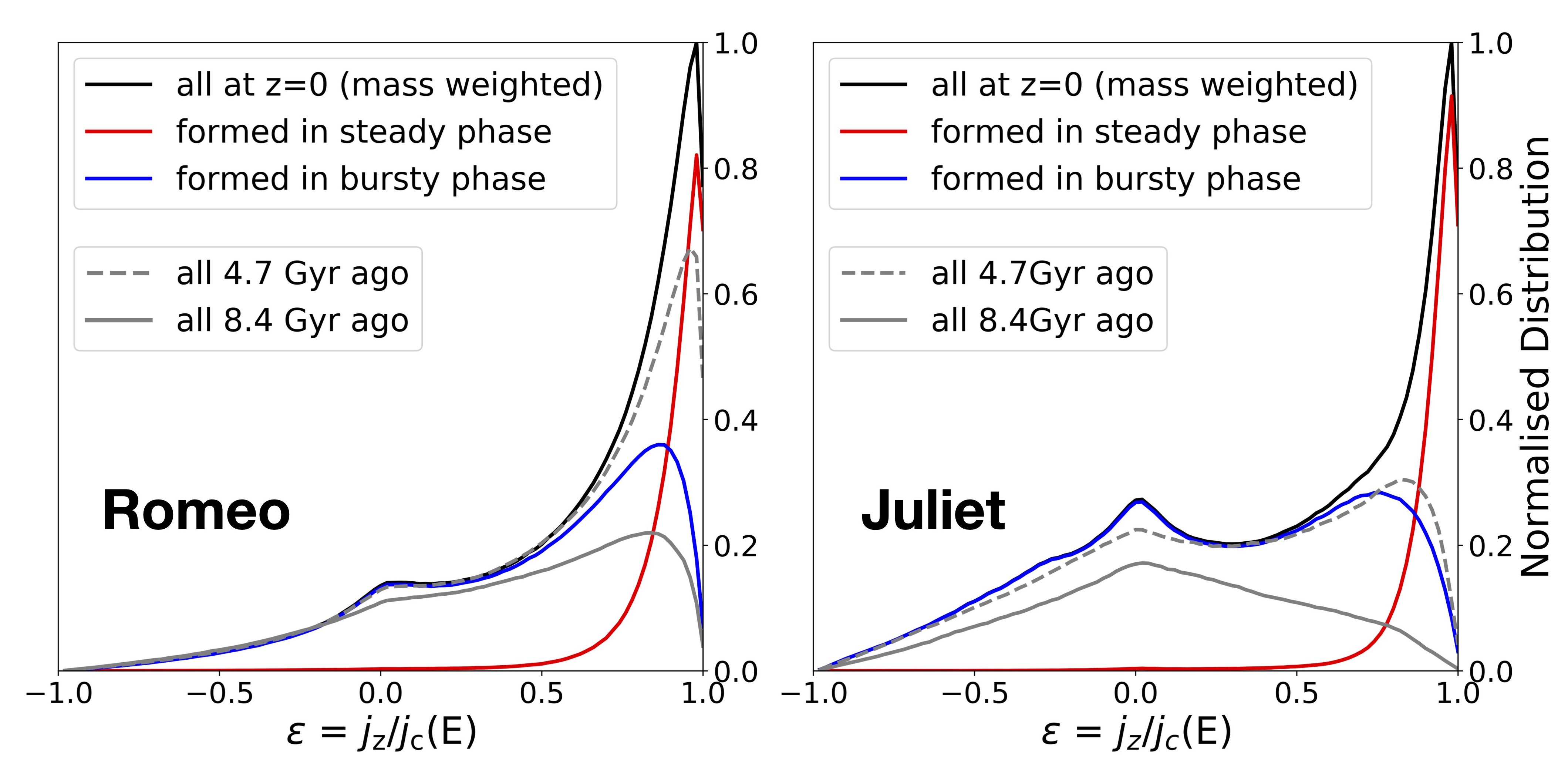}
    \caption[]{ Circularity distributions for stars in \Romeo~(left) and \Juliet~(right). The solid black lines in each case show the mass-weighted distribution of stellar circularities ($\epsilon$) for all stars within $R_{90}$ of each galaxy ($13.3$ kpc and $9.6$ kpc, respectively) at $z = 0$. The solid red and blue lines correspond to stars formed in the steady star formation phase and the bursty phase, respectively. In each galaxy, stars that formed during the steady phase have high circularities, peaked near $\epsilon = 1$. Stars formed during the bursty phase are much less ordered, with a coherent rotation peak at $\epsilon \lesssim 0.8$, indicative of thick-disc kinematics.  The gray solid and dashed lines show {\em total} mass-weighted star-particle distributions for the main progenitors of each galaxy $8.4$ Gyr ago and $4.7$ Gyr ago, respectively.  \Romeo~(left), which ended its bursty phase 6.5 Gyr ago, had already developed a fairly substantial high-angular momentum peak $4.7$ Gyr ago (dashed line), whilst \Juliet~(right), which ended its bursty phase only 4.4 Gyr ago, had only a modest thick-disc like peak at the same time. Further back in time, 8.4 Gyr ago, \Romeo~ was much less ordered than it is today, but still had a thick-disc-like peak in stellar circularity.  \Juliet, on the other hand, had mostly spheroid-like kinematics, with a circularity distribution centred on $\epsilon \sim 0$, at this time.}
	\label{fig:circ_dist_260_400}
\end{figure*}

The 8.4 Gyr and 4.7 Gyr panels in Figure \ref{fig:Juliet_morph_compare} for \Juliet~show differences in morphology with time that are representative across our larger simulated sample. Specifically, we find that the bursty phase itself can be further divided into two periods of morphological development: 1) a very early, chaotic bursty phase, where even the youngest stars ($<50$ Myr) have angular momenta that are misaligned with the existing stars in the galaxy; and 2) a later, quasi-stable ``bursty-disc" phase, where some short-lived angular momentum cohesion exists.  As shown with an example in the next section,  we find that stars that are born very early on, when the SFR is {\em very} bursty, tend to be born with spheroidal-type orbits (with peaks in the $\epsilon$ distribution ranging from $0-0.3$).  Stars that are made during the later, quasi-stable bursty-disc phase, tend to be fairly coherent for a short period of time, with circularity distributions within $\sim$50 Myr of their birth that straddle thin/thick disc characteristics (peaking with $\epsilon \sim0.6-0.8$).  These bursty-disc stars are quickly heated to thicker-disc orbits within $\sim$100 Myr \citep[similar to the behavior reported by][]{Meng2020}. This later heating appears to be a result of bursty feedback and chaotic accretion. Similar components could also be found based on stellar populations at $z=0$ using a Gaussian mixture model (Nikakhtar et al., in preparation).

Because this paper focuses on thin- and thick-disc formation, we have refrained from presenting results on early {\textit{in-situ}} spheroid formation, though this would be an interesting topic for future work. It is worth noting that, when weighted by luminosity, the spheroidal components contribute minimally to the total light in our galaxies at $z=0$.  

\subsubsection{Kinematic classification with time}

Figure \ref{fig:circ_dist_260_400} shows mass-weighted circularity distributions for star particles in \Romeo~(left) and \Juliet~(right).   The black solid lines indicate distributions for {\em all} stars within $R_{90}$ of each galaxy at $z=0$. The blue lines indicate the $z=0$ circularities for stars that formed during the early, bursty phase ($t_{\rm birth} < t_{\rm B}$), whilst the red lines refer to those formed during the later steady phase. Stars that formed during the steady phase are much more circular (thin-disc like) in each case, peaking close to $\epsilon = 1$. The stars that formed during the early bursty phase show much less coherence in angular momentum, with high-$\epsilon$ peaks closer to $\epsilon \sim 0.8$, indicative of thick-disc kinematics. Note that the distributions are normalised such that the sum of the red and blue lines equals the black lines.

The gray curves in Figure \ref{fig:circ_dist_260_400} show the $\epsilon$ distributions for all stars in the main progenitor of each galaxy at two different lookback times: 8.4 Gyr ago (gray solid) and 4.7 Gyr ago (gray dashed).  These are the same times visualised in the lower left and lower middle panels of Figures \ref{fig:Romeo_morph_compare} and \ref{fig:Juliet_morph_compare}. \Romeo, which had just finished its bursty phase by $4.7$ Gyr ago, had a fairly prominent peak at high circularity at that time. \Juliet, which was still in its bursty phase at that time, had a less well-ordered angular momentum distribution. Both galaxies were systematically less well-ordered 8.4 Gyr ago than they were 4.7 Gyr ago. Whilst \Romeo~had a small peak near $\epsilon \sim 0.8$, more characteristic of a thick disc component,  \Juliet~had a distribution peaked near $\epsilon = 0$, as expected for a spheroidal system. These differences in angular momentum structure mirror the morphological differences between these two galaxies at the same times shown in Figures \ref{fig:Romeo_morph_compare} and \ref{fig:Juliet_morph_compare}. 

\begin{figure*}
	\includegraphics[width=0.95 \textwidth, trim = 0.0 0.0 0.0 0.0]{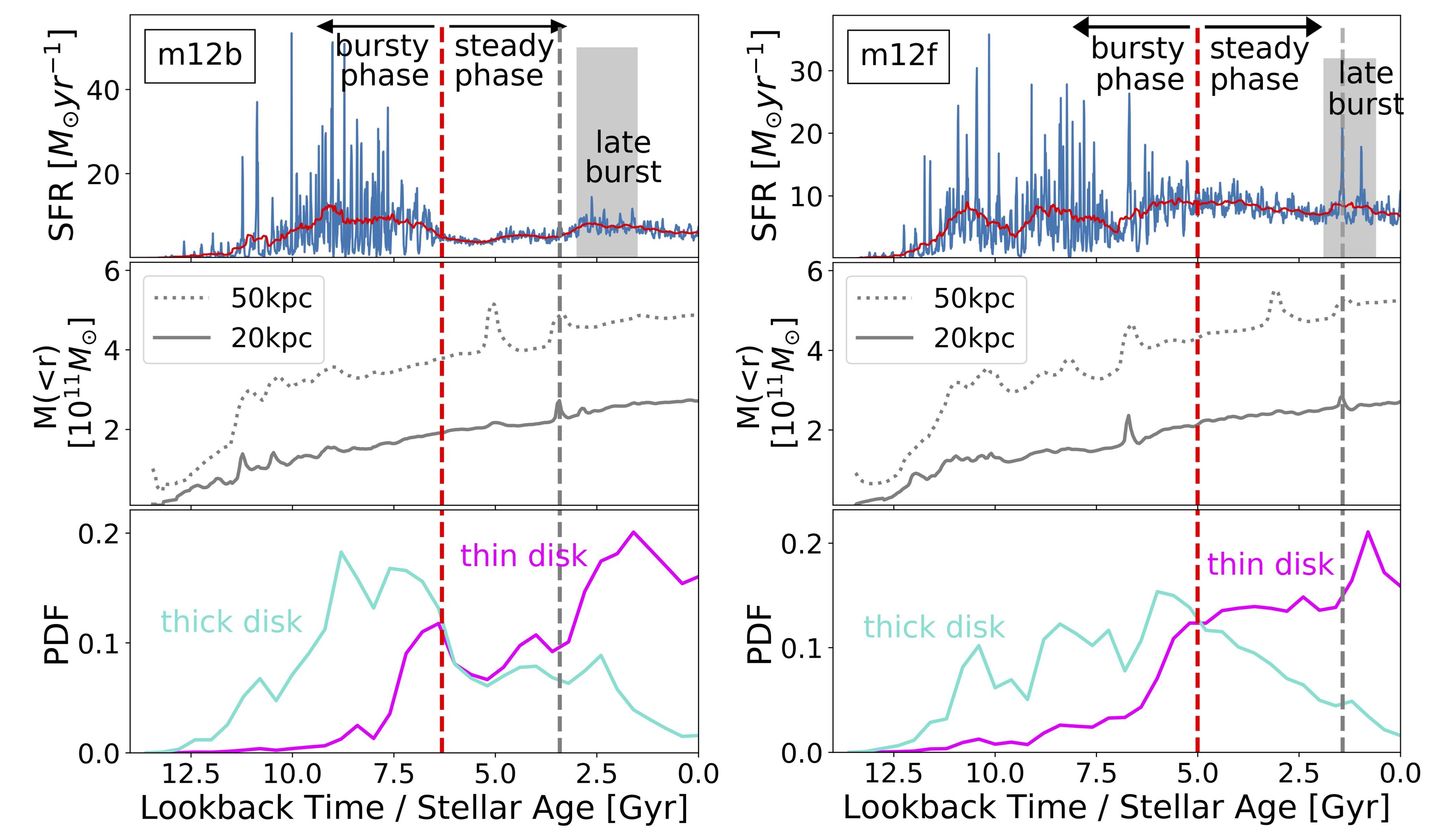}
    \caption[]{Star formation histories, mass accretion histories and thin/thick disc stellar age distributions for \texttt{m12b} (left) and \texttt{m12f} (right).  These are examples of galaxies with late-time bursts triggered by minor mergers (shaded gray bars).  The {\bf top panels} show the star formation rate in each galaxy as a function of lookback time. The blue lines show the ``instantaneous" star formation rated averaged over 10 Myr bins, while the red lines show the ``smoothed" star formation rate averaged over 500 Myr bins. There are two distinct phases in the star formation history for each galaxy, an early bursty phase and a late steady phase.  We divide the two at a time $t_{\rm B}$, which we define as the time when the variance in instantaneous star formation rate falls below $0.2$ times the smoothed star-formation rate.  This "bursty-phase lookback time" is marked by the vertical red dashed line in each upper panel. The gray bands indicate late-time bursts of star formation that occur during the steady phase. The {\bf middle panels} show the total mass of central galaxies within 50kpc (grey dotted) and 20kpc (grey solid), respectively, as a function of lookback time. From the mass accretion history, we see that the fairly significant burst in \texttt{m12f} was triggered by a late-time, prograde, LMC-size merger.  The smaller burst in \texttt{m12b} was also triggered by the final coalescence of a merger, of similar size, but this time on a polar orbit\protect\footnotemark. We record the first central impact time of this type of mergers and mark with grey dashed lines in the plot.  The {\bf bottom panels} show the age distribution of $z=0$ stars that belong to the thick disc (cyan) and thin disc (magenta) in each galaxy.  We see that thick disc stars have ages that track closely the bursty period of star formation while thin-disc stellar ages correspond more closely to the steady phase.  Stars made in the late-time bursts appear to populate the thin disc primarily, but some stars end up in the thick disc as well. 
    The burst age is proceeded by an enhanced tail of slightly older thick-disc stars, which is consistent with what would be expected from disc heating. These events do not change appreciably the median age of thick disc stars but do enhance the post-steady-phase tail of the thick-disc stellar age distributions compared to cases without late bursts (compare to Figure \ref{fig:SFH_Age_distribution_RJ}).
    }
	\label{fig:SFH_Age_distribution_bf}
\end{figure*}
\footnotetext{The fall-in directions are confirmed with the visualizations of the two simulations.}

\subsection{Late-time mergers and starbursts}
\label{sec:mergers}

Three of our twelve galaxies (\texttt{m12b}, \texttt{m12c}, and \texttt{m12f}) experience late-time mergers after the steady phase has commenced (see Appendix \ref{sec:appen} for details).  We define a merger to be an event that impacts the central galaxy (< 20 kpc) with a satellite that had a total mass (baryons and dark matter) greater than $5 \times 10^{10}\msun$ when it crossed the inner $50$ kpc. We record this as the merger time. Additionally, seven of our other galaxies have mergers of comparable sizes during the bursty phase, but these mergers do not correlate with disc properties in significant ways (see Appendix \ref{sec:appen}).

Figure \ref{fig:SFH_Age_distribution_bf} illustrates the star formation histories (top panels), total mass (dark matter plus baryons within $50$ kpc and $20$ kpc) evolution (middle panels), and disc component age distributions (bottom panels) associated with \texttt{m12b}~(left) and \texttt{m12f}~(right). The bottom panel splits the age distribution into thin (cyan) and thick (magenta) disc stars. \texttt{m12b} experiences a polar-orbit merger with a gas-rich, LMC-size satellite ($\sim 10^{9.5} \,$ \Msun in baryons, $\sim  10^{11}$ \Msun in dark matter) that coalesces at the time of the late-burst marked.   The more prominent late-burst in \texttt{m12f} is associated with a merger with a satellite of a similar mass, but this time on a prograde orbit.

These late-time mergers and associated bursts do not change broad correlations we find between bursty-phase lookback time and thin-disc fractions and {\em median} thick-disc ages. However, they do enhance the age distribution of the youngest thick-disc stars. The lower panels of Figure \ref{fig:SFH_Age_distribution_bf} include examples of this effect, where the thick-disc age distributions are not as sharply truncated after the bursty phase as they are in Figure \ref{fig:SFH_Age_distribution_RJ}. This seems mostly to arise from heating associated with the merger, but feedback from the burst could also contribute. Interestingly, the burst also coincides with a peak in the {\em thin-disc} stellar age distribution.  Many of the stars that form in these bursts apparently retain thin-disc orbits. That gas-rich mergers can promote stellar-disc formation is a well-known phenomenon \citep{Robertson06}.  \citet{Santistevan2021} use the same simulations we analyze here to show that existing metal-poor stars and low-metallicity gas deposited in LMC-size mergers can explain the existence of low-metallicity prograde stars in the Milky Way \citep{Sestito20}.
 
One of our twelve galaxies (\texttt{Thelma}) experiences a late-time burst ($\sim$1 Gyr lookback time) that is {\em not} associated with a merger.  This appears to be a stochastic event associated with the fact that \texttt{Thelma}~ has only recently settled down to  $\sigma_{10}/{\rm SFR}_{500} < 0.2$ at $t_{\rm B} = 2.6$ Gyr. Unlike the majority of our galaxies, \texttt{Thelma} does not settle down to a variance much smaller than $0.2$; so the ``burst'' by our formal definition looks more like a stochastic event. Only one other galaxy in ours sample, \texttt{m12w}, never really settles down ($t_{\rm B} = 0.0$ Gyr) -- its  variance in instantaneous SFR around $z=0$ is still $\sim 0.3$.  

\subsection{Sample-wide trends}
\label{sec:sample-wide}

\begin{figure*}
	\includegraphics[width=0.98 \textwidth, trim = 0.0 0.0 0.0 0.0]{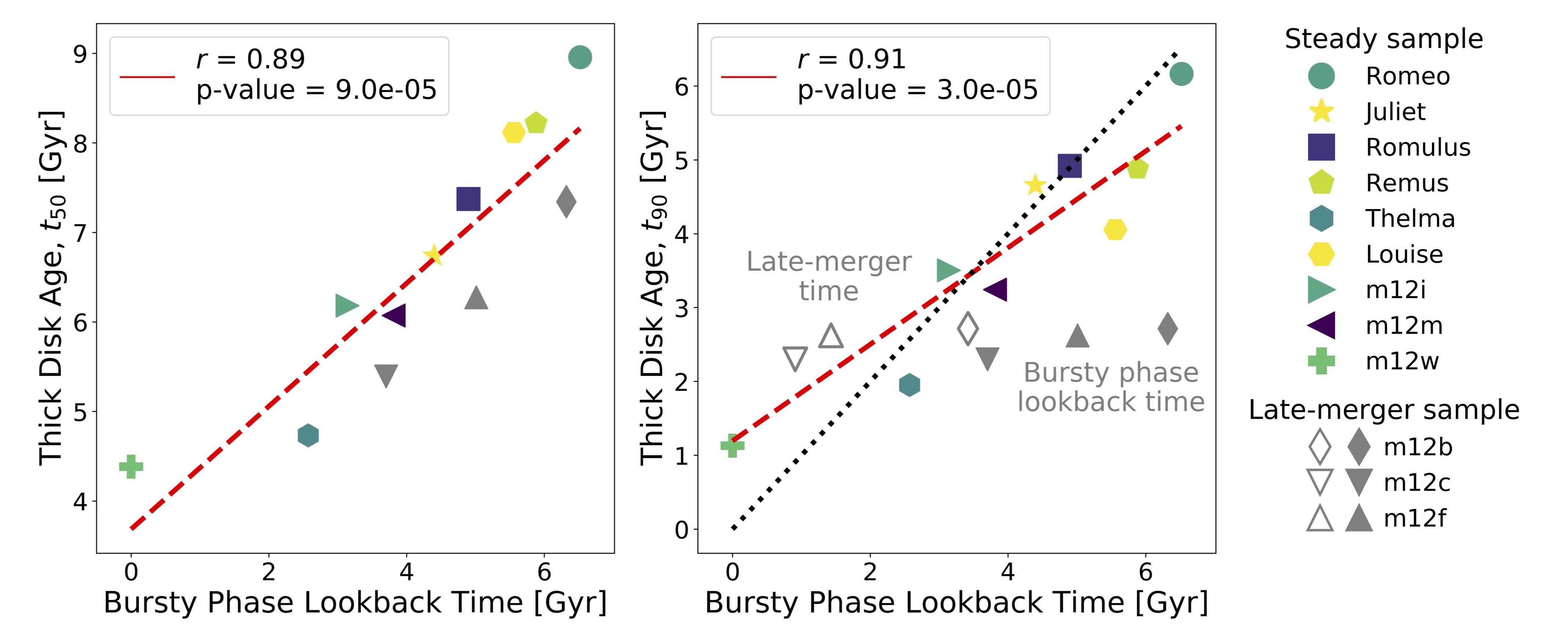}
    \caption[]{ 
    Correlation between the bursty-phase lookback time and thick disc age for our entire sample.  The gray points correspond to galaxies that experience late-time minor mergers after the steady phase has begun. {\bf Left:}. Median thick-disc age ($t_{50}$) versus the lookback time to the end of the bursty phase. The legend (far right) displays a unique symbol type per galaxy. The dashed-red line shows a linear fit to the relation. The Pearson correlation coefficient ($r = 0.89$) is boxed in the upper left. The typical age of a thick-disc star correlates quite strongly with the end of the bursty phase. {\bf Right:} Ninetieth percentile oldest thick-disc star age ($t_{90}$) versus the lookback time to the end of the bursty phase (solid points).  The solid gray points -- those with late-time mergers -- clearly lie on a different relation than the coloured points, suggesting that the late bursts influence and populate the young-star tail of the thick disc population.  The open points are the same galaxies, now depicted at the lookback time when the late-time merger occurred. These seem to align fairly closely to a one-to-one line (gray dotted), along with the coloured points. The red dashed line shows a linear fit to the open gray and coloured points.
     }
	\label{fig:bursty_time_thick_age}
\end{figure*}

\begin{figure*}
	\includegraphics[width=0.95 \textwidth, trim = 80.0 20.0 100.0 20.0]{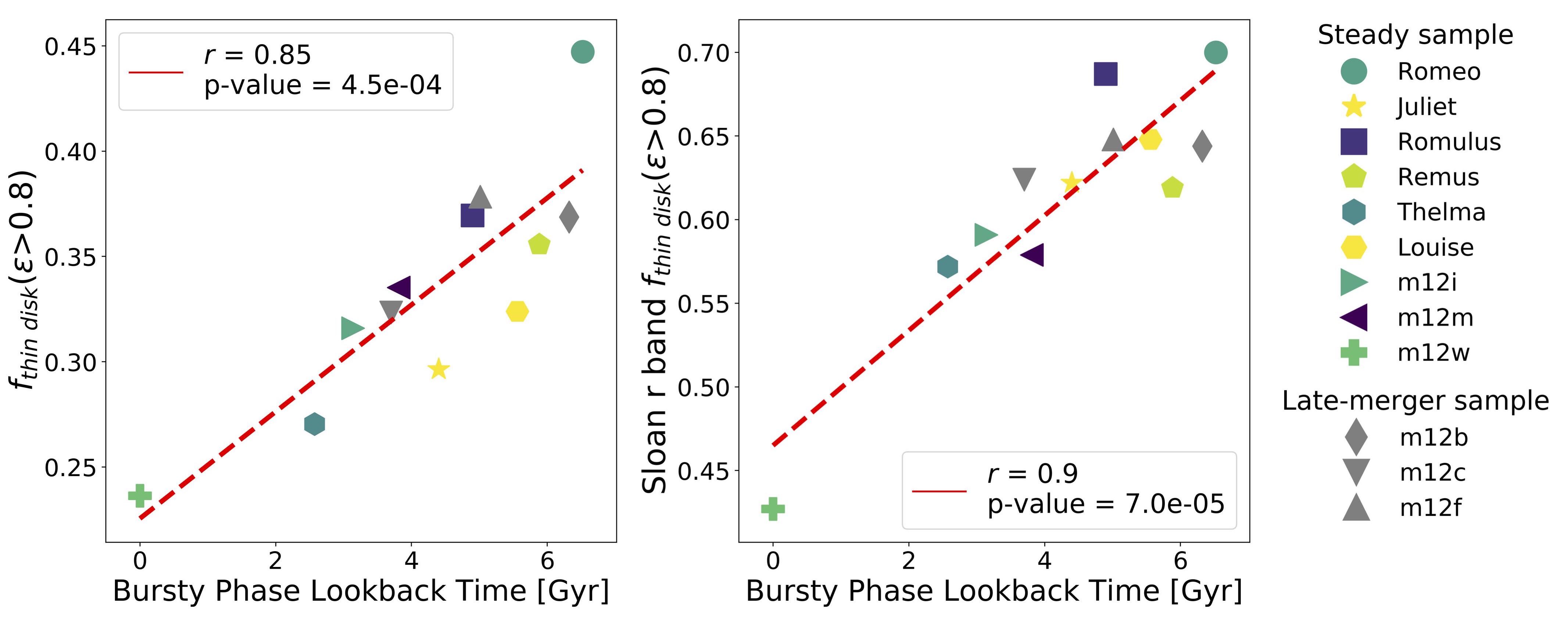}
    \caption[]{Correlations between the bursty-phase lookback time and the thin-disc fraction. {\bf Left:} Mass-weighted thin-disc fraction versus bursty-phase lookback time for each run. The legend on the far-right relates each symbol to a unique galaxy in our sample, in the same manner as Figure \ref{fig:bursty_time_thick_age}.  {\bf Right:} Luminosity-weighted thin-disc fraction versus bursty-phase lookback time.   In both panels, galaxies with longer lookback times to the bursty phases (and hence the longer-lasting steady phases) have more pronounced thin disc. The red-dashed line in each panel shows a linear fit to the points. The corresponding Pearson correlation coefficient is listed. The correlation is significant ($r > 0.8$) in each case, even though we have included no information on the relative average rate of star formation in the bursty (thick-disc) phase compared to the steady (thin-disc) phase.  As in Figure \ref{fig:bursty_time_thick_age}, we present the runs that have a recent minor mergers in gray. There is no significant difference between these runs and the others in the thin-disc fraction at fixed bursty-phase lookback time.
    }
	\label{fig:bursty_time_relation}
\end{figure*}

\begin{figure*}
	\includegraphics[width=0.98 \textwidth, trim = 0.0 0.0 0.0 0.0]{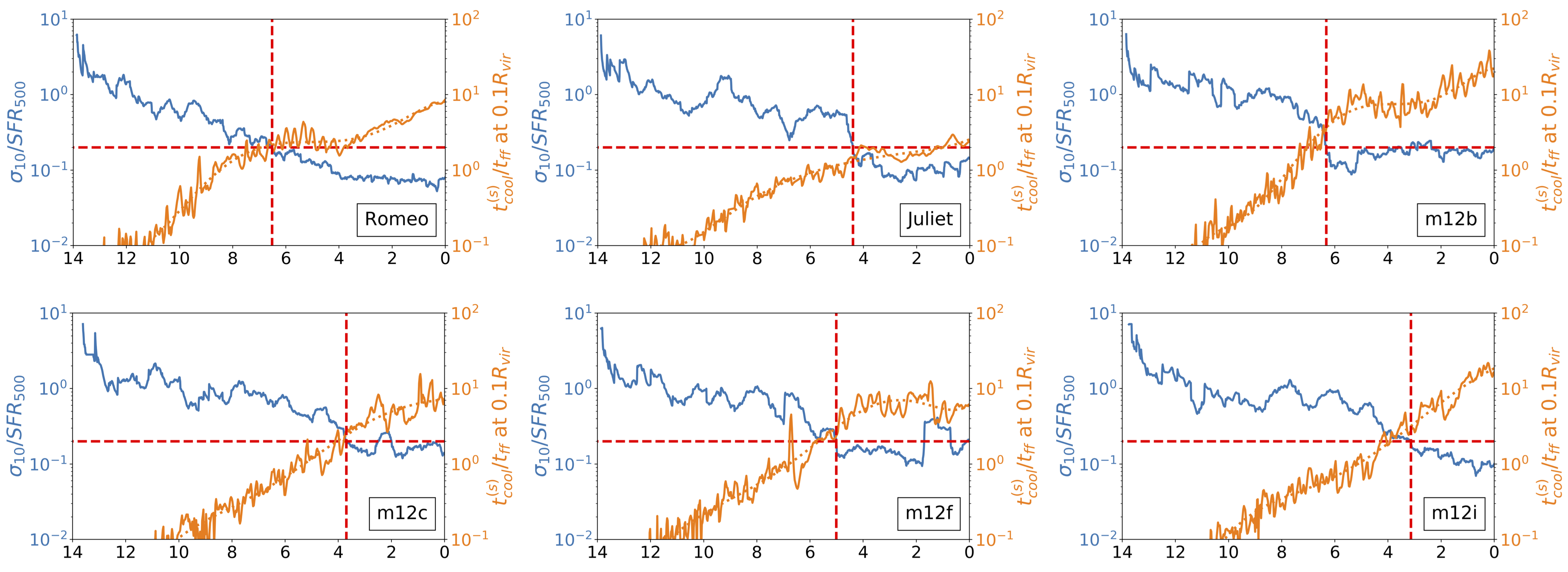}
    \caption[]{Parameters that track star-formation burstiness (blue) and inner CGM virialisation propensity (orange) as functions of lookback time for six of our haloes.  As described in 3.3.3, the orange lines (right axis) show the cooling time to free-fall time ratio measured at $0.1R_{\mathrm{vir}}$ as a function of lookback time. The blue lines (left axis) shows the variance in instantaneous star formation rate divided by the smoothed star formation rate as a function of time. The  horizontal red dashed line marks the threshold above which the inner CGM virialises, $\tcoolsh/\tff\approx2.0$, as defined in \citet{Stern20}.  The axes are set so that the same line corresponds to $\sigma_{10}/$SFR$_{500}= 0.2$, which we have adopted in this paper to define the end of the bursty phase. The bursty-phase lookback time $t_{\rm B}$ defined in this paper is marked by the vertical red dashed line.  Note that the same time roughly corresponds to the time when the inner CGM becomes virialised.
    }
	\label{fig:tcooltff_sigmamean}
\end{figure*}

\subsubsection{Thick disc age}
\label{sec:age-trends}

Using four illustrative examples, Figures \ref{fig:SFH_Age_distribution_RJ} and \ref{fig:SFH_Age_distribution_bf} suggest that the ages of kinematically-identified thick-disc stars at $z=0$ tend to track the period of bursty star formation in these galaxies. The left panel of
Figure \ref{fig:bursty_time_thick_age} demonstrates that these trends hold for our entire sample.  Displayed is the median age of thick-disc stars ($t_{50}$) versus the bursty-phase lookback time ($t_{\rm B}$) for each galaxy.  The correlation is quite tight, with more recent bursty phases associated with younger thick-disc ages.  Note that $t_{\rm B}$ along the horizontal axis is determined entirely from the star formation history of the galaxy and includes no dynamical information whatsoever, and thus the observed correlation is nontrivial. For example, if thick discs were formed primarily from initially thin discs that were heated by mergers, we would expect no such correlation.   

The typical (median) thick-disc star was formed approximately 3 Gyr prior to the end of the bursty phase. The red-dashed line shows the best-fit linear relation:
\begin{equation}
    t_{50} = 3.7 + 0.69 t_{\rm B}, 
\end{equation}
where times are assumed to be in units of Gyr. The Pearson correlation coefficient for the points in the left panel is $r=0.89$ with p-value = $9.0 \times 10^{-5}$. Although not shown, we find that that the average age of thick disc stars produces a very similar trend with bursty-phase lookback time as the median age displayed here.  Given that the thick-disc population is primarily born during the bursty phase, it is natural to ask if the youngest thick-disc stars allow us to age-date the end of the bursty phase in a one-to-one way. We find that this is true only for the nine of our twelve galaxies that {\em do not} have a late-time merger during the steady phase.

The right panel of Figure \ref{fig:bursty_time_thick_age} shows the age of the ninetieth percentile oldest thick-disc star ($t_{90}$) versus the bursty-phase lookback time (solid points). The dotted gray line shows the one-to-one relation for reference.  The gray symbols refer to galaxies with late-time mergers, which clearly deviate from the trend. The one galaxy in our sample that experiences a late-time burst  not triggered by a merger (\texttt{Thelma}, coloured pentagon) does not deviate significantly. The open gray symbols use the lookback time to the late-time merger as the horizontal coordinate.  With this choice, the points fall along a fairly tight relation (with Pearson correlation coefficient of $r=0.91$ and p-value = $3.0 \times 10^{-5}$). The dashed red line shows a linear fit to the coloured and open points (with solid gray points ignored):
\begin{equation}
    t_{90} = 1.2 + 0.65 t_{\rm B}, 
\end{equation}
where times are assumed to be in unites of Gyr.

The gray points in the left panel of Figure \ref{fig:bursty_time_thick_age} represent galaxies with late-time mergers. This group tends to track the relation, but also tend to lie systematically below the average trend with respect to median age.  This is consistent with the interpretation that the young-star tail of the thick-disc population has been populated by stars formed after the end of the bursty phase. Nevertheless, the fraction of stars populated in this way is small enough ($\lesssim 10\%$) that the broad trend with median age and bursty phase lookback time is preserved. 

Given that the youngest thick-disc stars may be associated with either the end of the bursty phase or a late-time merger, it maybe be difficult to use the age of the youngest stars to easily date the end of the bursty phase.  In principle, one could look for features in the age distribution of thick-disc stars to gain insight on these questions (see Figure \ref{fig:SFH_Age_distribution_bf} where the bursty lookback time does seem to imprint a feature in the age distribution of thick-disc stars).  However, it will likely be more straightforward to use the typical age (median or average) of thick-disc stars to estimate the lookback time corresponding to the end of the bursty phase and the beginning of the steady phase (independent of the recent merger activities).

\subsubsection{Thin disc fraction}
\label{sec:fraction-trends}

Figure \ref{fig:bursty_time_relation} shows the correlation between bursty-phase lookback time and thin-disc fraction, with each symbol type mapped to a specific galaxy name (far-right legend). The left panel employs a mass-weighted thin-disc fraction, whilst the right panel uses as luminosity-weighted thin-disc fraction. The red lines show linear fits to the data points. In each case, the correlation is strong, but with scatter, with Pearson correlation coefficients of $r=0.86$ (mass weighted) and $r=0.88$ (luminosity weighted). Both p-values ($3.9 \times 10^{-4}$ and $1.6 \times 10^{-4}$, respectively) are much less than the significance level. 

We see that the earlier the lookback time to the transition, the more prominent the thin disc is. It makes qualitative sense that the longer stars are created in the ``settled'' phase, the larger the fraction of thin disc stars we would see.  
At fixed thin-disc fraction, we see $2-3$ Gyr scatter in the lookback time to the bursty phase. It would be surprising, however, if this relation were any tighter, as it contains no information on the absolute star formation rates in either phase.  Specifically, at fixed lookback time to the transition, the higher the average star formation rate during the thin-disc/steady phase compared to the thick-disc/bursty phase, the more prominent the thin disc would be. We see that this trend generally holds for our galaxies. For example, if we examine the star formation histories in Figure \ref{fig:SFH_Age_distribution_bf} for galaxies \texttt{m12b}~ and \texttt{m12f}, we see that \texttt{m12b} has a higher smoothed-average star formation rate during the bursty phase than it does during the steady phase. Conversely, \texttt{m12f} has a similar smoothed-average star formation rate before and after the transition. This means that \texttt{m12b} will be making fewer thin-disc stars per unit time during the steady phase than  \texttt{m12f}. This explains why \texttt{m12b} has a thin-disc fraction (0.64 in luminosity) that is slightly {\em lower} than \texttt{m12f} (0.65), even though its steady phase lasts more than one billion years longer (7.34 Gyr vs. 6.28 Gyr).  

\subsubsection{CGM virialisation and steady star formation}
\label{sec:CGM-trends}
The physical origin of the progression from early, bursty and less kinematically-ordered star formation to late-time, steady star formation in thin discs is not clear.  An important clue comes from the work of \citet{Stern20}, who used  FIRE-2 simulations to show that the bursty to steady transition in galaxy star formation coincides with virialisation of the inner CGM. They quantify inner CGM virialisation using the ratio of the cooling time of shocked gas $\tcoolsh$ to the free-fall time $\tff$ at an inner radius $r = 0.1 R_{\rm vir}$.  
When $\tcoolsh / \tff \gtrsim 1$ the inner CGM is smooth and largely supported by thermal pressure. In contrast, when $\tcoolsh / \tff \lesssim 1$, the inner CGM has large pressure fluctuations and is highly dynamic.
Using a sample of sixteen zoom simulations with halo masses ranging from M$_{\rm halo}$ $=10^{10.6} - 10^{13}$ M$_\odot$, \citet{Stern20} shows that gaseous discs become rotationally supported and star formation transitions from bursty to steady at roughly the time when the ratio first becomes $\tcoolsh / \tff > 2$.  Their sample included four of the twelve galaxies we consider here.

In this brief subsection, we extend the \citet{Stern20} analysis to the additional haloes in our sample and confirm their reported trends.  Using the same definitions of free-fall time and cooling time described in section 2.1 of their paper, we show that inner CGM virialisation at the time when $\tcoolsh / \tff = 2$ generally coincides with our bursty to steady SFR transition at $\sigma_{10}/$SFR$_{500}= 0.2$ (Eq. \ref{eq:def}).  This is demonstrated in Figure \ref{fig:tcooltff_sigmamean}, where we show the evolution of the inner virialisation parameter (right axis, orange line) and the burstiness parameter (left axis, blue line) as functions of lookback time for six example haloes.  Not only do the transition timescales coincide in each case, but the monotonic progressions of each parameter tends to evolve inversely with the other in time.  At early times, when gas flows are prone to cooling instabilities and clumpy accretion, the star formation is more bursty.  At late times, when cooling times are long and the flow can be relatively smooth and well-mixed, star formation tends to be more constant.  We find similar behaviors hold for every halo in our sample.

The proceeding analysis demonstrates that as the inner CGM of our galaxies virialises, the star formation becomes less bursty (Fig. \ref{fig:tcooltff_sigmamean}).  This is also the time when stars tend to be formed with thin-disc kinematics (Fig. \ref{fig:bursty_time_thick_age}). One hypothesis that explains this, suggested by \citet{Stern20}, is that a virialised inner CGM enables the formation of stable discs because a hot and uniform halo can pressure-confine disruptive  superbubbles driven by clustered supernovae.  Another possibility is that smooth and well-mixed accretion enables more coherently aligned angular momentum at the time of accretion onto the galaxy (Hafen et al., in preparation).  These issues are important topics for further exploration.


\section{Discussion and Conclusions}
\label{sec:conclusions}

We investigate the formation of the stellar thin and thick disc components using twelve FIRE-2 zoom-in simulations of Milky-Way-mass galaxies. Our main findings include the following:
\begin{itemize}

\item Each galaxy experiences an early period of bursty star formation that transitions into a steady phase, with a relatively constant star formation rate at late times.  The transition time corresponds to the time when the inner CGM becomes sub-sonic and virialises (Figure \ref{fig:tcooltff_sigmamean}). 

\item The transition from bursty to steady star formation correlates with a shift in the formation of stars with thick-disc kinematics  to thin-disc kinematics (Figures \ref{fig:SFH_Age_distribution_RJ}, \ref{fig:Romeo_morph_compare}, and \ref{fig:Juliet_morph_compare}).

\item The lookback time to the end of bursty phase ranges from $t_{\rm B} = 0.0 - 6.5$ Gyr in our sample.  This time correlates strongly with the median age of thick disc stars at $z=0$ (Figure \ref{fig:bursty_time_thick_age}). 

\item Galaxies with longer steady phases (larger $t_{\rm B}$) tend to have higher thin-disc fractions (Figure \ref{fig:bursty_time_relation}).

\end{itemize}

Three of our twelve simulations have appreciable late-time mergers that occur after the steady (thin-disc) phase has commenced. These mergers are not responsible for the bulk of thick-disc stars, though they do heat some disc stars and populate the young-star tail of the thick disc population demonstrated in Figure \ref{fig:SFH_Age_distribution_bf} and the right panel of Figure \ref{fig:bursty_time_thick_age}.

The fact that our discs emerge thick and become thinner over cosmic time is consistent with previous findings of ``upside-down'' disc formation \citep{Brook04,Brook12,Bird13,Navarro18,Bird20,Ma2017,Park20}. However, our result that the transition is associated with a transition in star formation activity -- from bursty to steady -- adds a new element to this discussion. That FIRE simulations of Milky-Way-mass haloes experience such a transition in star formation activity is not a new result \citep{Muratov15,Sparre17,AA17,FG18}.  The onset of the steady phase appears to be related to the virialisation of the inner CGM \citep[][and Figure \ref{fig:tcooltff_sigmamean}]{Stern20}.  A hot, pressurised CGM may stabilise the disc against supernovae-driven outflows and enable thin-disc formation \citep{Stern20}. If correct, this interpretation opens the possibility of using stellar archaeology to learn about the origin of the Milky Way's CGM and its associated history of star-formation modes.

Whilst a more observationally-oriented comparison is required to interpret our results for the Milky Way confidently, it is tempting to explore some potential implications based on na\"{i}ve comparisons to published estimates of the Galactic thick-disc age distribution \citep{Haywood13,Snaith14,Martig16,Hayden17,Sharma19}. Most estimates suggest that the Milky Way thick-disc has a median age of $\sim 9$ Gyr, with few stars younger than $6$ Gyr. Such an age distribution is most similar to our \Romeo~simulation (Figure \ref{fig:SFH_Age_distribution_RJ}), which transitioned from bursty to steady star formation $\sim 6.5$ Gyr ago. This simple comparison would suggest that the Milky Way transitioned at a similar time, commensurate with the virialisation of its inner hot halo. If this is the case, then prior to that time, the Milky Way lacked a dominant thin disc component, was forming stars in a bursty manner, and had non-virialised inner CGM.

Related to our analysis, \citet{Bellardini2021}, using the same set of simulations, finds that gas disc metallicity in-homogeneity was dominated by azimuthal variations at high redshift but then transitioned to being dominated by radial gradients at lower redshifts, which has also been reported in the previous analysis in the FIRE-1 simulations across a much wider galaxy mass range \citep{Ma2017_2}. The transition epochs after which radial variations dominate over azimuthal scatter agree broadly well with our measurement of the transitions from bursty to steady phase. Although there is significant scatter and some time delay between the measurements of two transition times, it shows some potential observable implications of the bursty/steady transition for galactic archaeology.

Some cosmological galaxy formation simulations, including those that demonstrate upside-down disc formation \citep[e.g.][]{Park20}, do not have early bursty star formation phases of the kind we witness in our models. This may partially be due to star formation threshold employed. Burstiness is suppressed in simulations with modest threshold densities ($\sim10$ cm$^{-3}$) for star formation, whereas our simulations require $\sim1000$ cm$^{-3}$  \citep[see][for a related discussion]{BL19,Dutton2019}. Given this, chemical tracers among the various Galactic kinematic components may provide a means to test star formation prescriptions. Other factors like the ISM model, local star formation efficiency and the stellar feedback model might also play a role. Future work in this direction will be illuminating.

\section*{Acknowledgements}
SY and JSB were supported by NSF grants AST-1910346 and AST-1518291. 
CK was supported by a National Science Foundation Graduate Research Fellowship Program under grant DGE-1839285.  
JS is supported also by the German Science Foundation via DIP grant STE 1869/2-1 GE 625/17-1 
AW received support from NASA through ATP grants 80NSSC18K1097 and 80NSSC20K0513; HST grants GO-14734, AR-15057, AR-15809, and GO-15902 from STScI; a Scialog Award from the Heising-Simons Foundation; and a Hellman Fellowship.
Support for JM is provided by the NSF (AST Award Number 1516374).
ZH was supported by a Gary A. McCue postdoctoral fellowship at UC Irvine.
ABG was supported by an NSFGRFP under grant DGE-1842165 and was additionally supported by NSF grants DGE-0948017 and DGE-145000.
Support for PFH was provided by NSF Research Grants 1911233 \&\ 20009234, NSF CAREER grant 1455342, NASA grants 80NSSC18K0562, HST-AR-15800.001-A. Numerical calculations were run on the Caltech compute cluster ``Wheeler,'' allocations FTA-Hopkins/AST20016 supported by the NSF and TACC, and NASA HEC SMD-16-7592.
CAFG was supported by NSF through grants AST-1715216 and CAREER award AST-1652522; by NASA through grant 17-ATP17-0067; by STScI through grant HST-AR-16124.001-A; and by a Cottrell Scholar Award and a Scialog Award from the Research Corporation for Science Advancement.
RF acknowledges financial support from the Swiss National Science Foundation (grant no 157591 and 194814).
We ran simulations using: XSEDE, supported by NSF grant ACI-1548562; Blue Waters, supported by the NSF; Pleiades, via the NASA HEC program through the NAS Division at Ames Research Center.




\bibliographystyle{mnras}
\bibliography{ref} 




\appendix
\section{Merger histories}
\label{sec:appen}

As discussed in Section \ref{sec:mergers}, we have explored the importance of  mergers in shaping thin/thick disc formation in our simulations.  In particular, we focus on merging events that perturb the total mass content within $20$ kpc by more than $2 \times 10^{10} \msun$ in the final coalescence of a satellite and that the merging satellite crossed within the $50$ kpc sphere for the first time with more than $5 \times 10^{10} \msun$ of total mass (which includes both dark matter and baryons).   We have made these choices because only above these thresholds do we discern any correlated activity that influences star formation or disc structure.  

Figure \ref{fig:SFH_Age_distribution_bf} showed the mass accretion histories (middle panels) for two galaxies that experience late-time mergers by this definition.  These mergers happened after the steady phase has commenced and appear to trigger late-time starbursts and also to add to the young-start tail of the age distribution of thick-disc stars. Figure \ref{fig:Courney_time_relation} shows example mass growth histories for two galaxies without such mergers. They also show star formation histories, specifically for \Romeo~ (left) and \Juliet~ (right).  The lower panels show total mass within $50$ kpc (dotted) and 20 kpc (solid) for each galaxy.  We see that the central galaxies themselves experience little in the way of merger activity that perturbs their overall masses going back prior to the time the steady phase began (red dashed lines). Nevertheless the transition from bursty to steady phase is sharp, and these transition times correlate with thick-disc ages (Fig. \ref{fig:bursty_time_thick_age}) and thin-disc fractions (Fig. \ref{fig:bursty_time_relation}).  

\begin{figure*}
\includegraphics[width=0.98\textwidth,trim = 0.0 0.0 0.0 0.0]{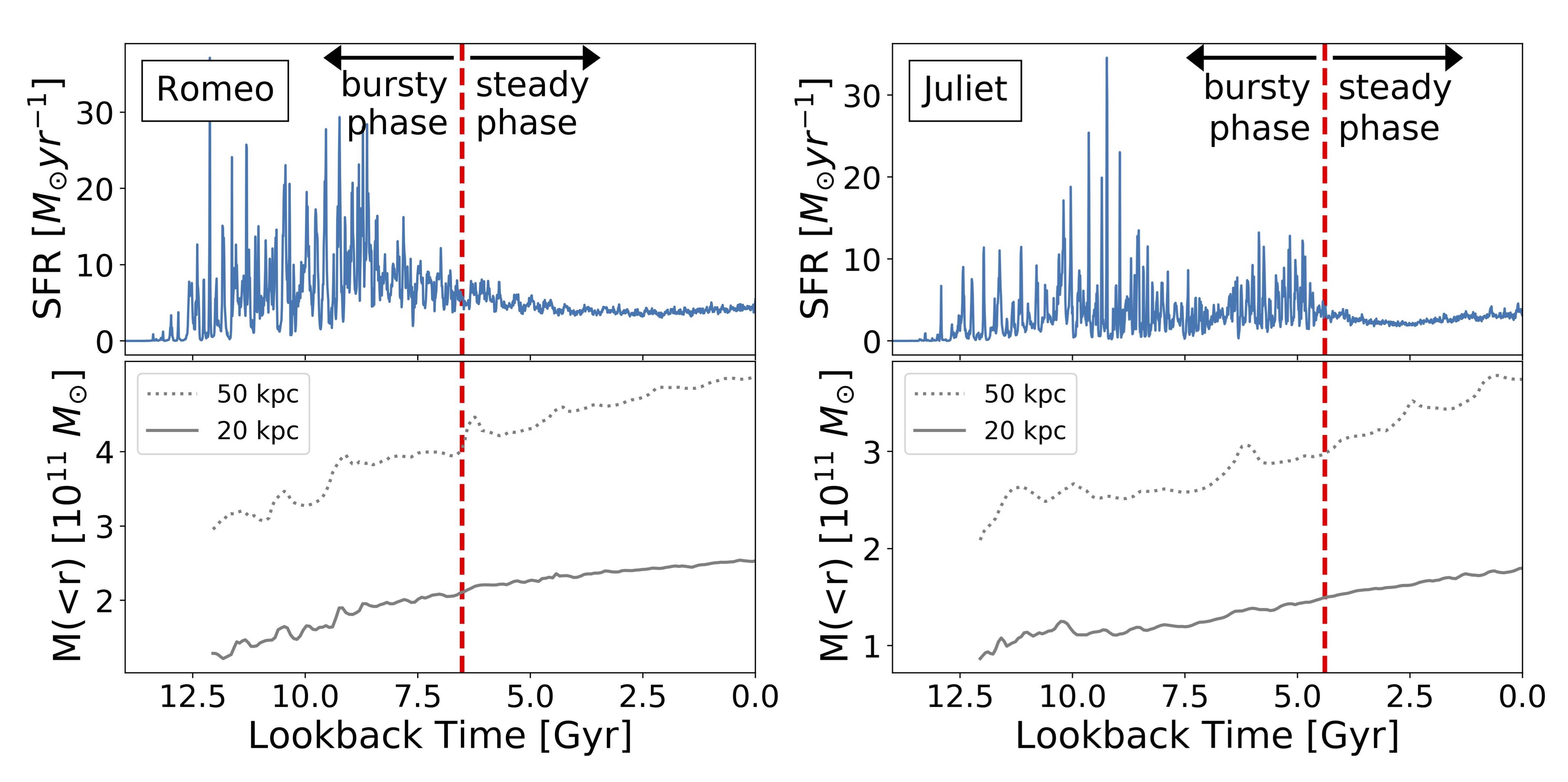}
\caption[]{Star formation histories and mass accretion histories for \texttt{Romeo} (left) and \texttt{Juliet} (right).  The solid and dotted lines in the lower panel show the total mass within 20 kpc and 50 kpc of the main galaxy, respectively, as a function of time.  We see that the mass growth is quite steady within 20 kpc, as expected for galaxies without significant merger activities.}
\label{fig:Courtney_data}
\end{figure*}

\begin{figure*}
	\includegraphics[width=0.95 \textwidth, trim = 50.0 0.0 50.0 0.0]{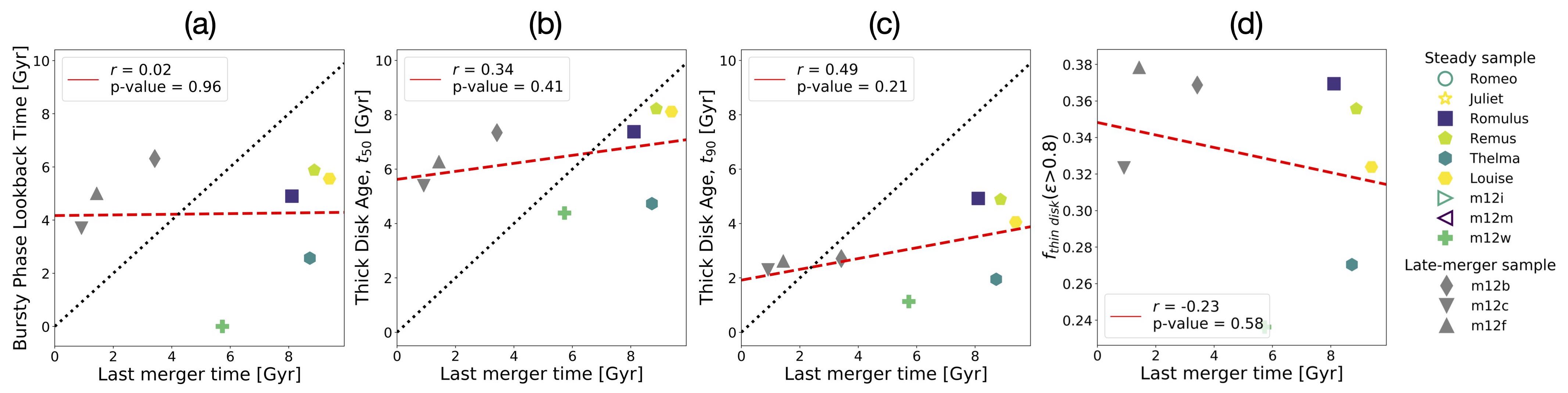}
    \caption[]{Correlations between the bursty-phase lookback time, thick-disc age, thin-disc fraction and the last merger lookback time. (a) Bursty-phase lookback time versus the last merger lookback time for each run. We record the merger time as the first central impact time as indicated by the grey dashed line in the middle panel of Figure \ref{fig:SFH_Age_distribution_bf}.  (b) Median thick-disc age ($t_{50}$) versus the last merger lookback time. (c) Ninetieth percentile oldest thick-disc star age ($t_{90}$) versus the last merger lookback time. (d) Mass-weighted thin-disc fraction versus the last merger lookback time. The legend (far right) shows the unique symbol type per galaxy similar to Figure \ref{fig:bursty_time_thick_age} and Figure \ref{fig:bursty_time_relation}. Note that, colored open markers are not plotted and they correspond to the galaxies that we do not identify any recent mergers from the mass accretion histories. In all panels, we see no significant correlations. }
	\label{fig:Courney_time_relation}
\end{figure*}

We have tabulated the lookback times of the last $M_{\rm tot} > \times 10^{10} \msun$  mergers in each of our simulated galaxies.   Figure \ref{fig:Courtney_data} shows these times plotted against the bursty-phase lookback time, thick disc age, and thin-disc fraction for our galaxies.  Four galaxies, \Romeo , \Juliet, \texttt{m12i}, and \texttt{m12m}, experience no such merger over their lifetimes and are not plotted.  We find no correlation between the last merger time and bursty-phase lookback time (left panel).  We find, at best, weak correlations with thick disc age. A more detailed analysis of possible correlations between mergers and bursty/steady transition is deferred for future work. 

\bsp	
\label{lastpage}
\end{document}